\documentclass[12pt]{iopart}

\usepackage{iopams} 
\usepackage{amssymb, wasysym}
\usepackage[pdftex]{graphicx}
\usepackage{color}
\usepackage{tabularx}
\newcolumntype{C}[1]{>{\centering\arraybackslash}p{#1}}

\usepackage{longtable}		
\usepackage{marvosym}
\usepackage{multirow}
\usepackage{multicol}
\usepackage{booktabs}

\usepackage{bbm}
\usepackage{rotating}
\newcolumntype{C}[1]{>{\centering\arraybackslash}p{#1}}

\usepackage{longtable}		
\usepackage{marvosym}
\usepackage{array}
\usepackage{verbatim}

\definecolor{orange}{rgb}{1,0.65,0}

\usepackage{pifont}
\usepackage{ifsym}
\usepackage{textcomp}
\usepackage{units} 
\usepackage{caption}[2008/08/24]

\usepackage[english]{babel}

\usepackage[hyphens]{url} %trennt URLs, wegen hyphens auch bei bindestrichen

\usepackage{lmodern} 
\usepackage[T1]{fontenc}

\definecolor{orange}{rgb}{1,0.65,0}

\newcommand{\text}[1]{#1}
\newcommand{\ce}[1]{#1}
\newcommand{\eqref}[1]{(\ref{#1})}
\newcommand{\V}{Vy\-cor\ }
\newcommand{\Vend}{Vy\-cor}
\newcommand{\degr}{$^{\circ}$}
\newcommand{\dC}{\,$^{\circ}$C}

\newcommand{\subs}[2]{#1_{\mbox{{\scriptsize #2}}}}

\renewcommand{\etal}{{\it et al.\ }}

\newcommand{\bild}[1]{Fig.~\ref{#1}}
\newcommand{\tab}[1]{Tab.~\ref{#1}}
\newcommand{\rel}[1]{Eq.~\eqref{#1}}

\begin{document}

\title[Capillary Rise and Imbibition of Liquids in Nanoporous Matrices]{Capillary Rise and Imbibition of Liquids in Nanoporous Matrices: Rheological Concepts and Experiments}

\author{Simon Gruener and Patrick Huber}
\address{Experimental Physics, Saarland University, D-66041 Saarbr\"{u}cken, Germany}
%\address{$^2$ Institute for Computer Science, Czestochowa University of Technology, Aleja Armii Krajowej 17, PL-42220 Czestochowa, Poland}
%\address{$^3$ Helmholtz Center for Materials and Energy, Glienicker Str. 100, D-14109 Berlin, Germany}
\eads{\mailto{s.gruener@mx.uni-saarland.de}, \mailto{p.huber@physik.uni-saarland.de}}

\begin{abstract}
Liquid flow propelled by capillary forces is one of the most important transport mechanisms in porous environments. It is governed by a fascinating interplay of interfacial, viscous drag as well as gravitational forces which liquids encounter upon invasion into geometries with often complex topologies, such as capillary networks of trees or interconnected fractures in soils and ice.  Whereas a detailed understanding of this phenomenon has been achieved on macroscopic scales, its phenomenology is poorly explored on the meso- and nanoscale, where predictions regarding its dynamics are hampered due to a possible break-down of continuum hydrodynamics and conflicting reports with respect to the conservation of the fluidity and capillarity of liquids upon spatial confinement. Here, we present fundamentals, concepts and an experimental, gravimetric study on the capillarity-driven invasion dynamics of liquids in networks of pores a few nanometers across in monolithic, nanoporous silica glass (porous Vycor). A variation of the complexity of the building blocks of the liquids investigated (water, chain-like n-alkanes, silicon oils and rod-like liquid crystals) along with a variation of the humidity and the temperature upon spontaneous imbibition allows us to gain information regarding the fluidity and capillarity of liquids in such nanoporous environments. We observe square-root of time imbibition dynamics for all liquids applied, which we can quantitatively describe by both a conserved bulk fluidity in the pore center and bulk capillarity at the advancing menisci, if we assume a sticky boundary layer (negative velocity slip length). Moreover, pecularities of nanopore-confined liquids, such as transport via the vapor phase leading to preadsorbed liquid layers, have to be properly accounted for. Upon increasing the chain-length in the case of the n-alkanes, we found hints towards a transition from stick- to slip-flow at the pore walls with increasing chain-length and thus polymeric behavior. Meniscus freezing, that is the formation of a surfactant-like, rectified monolayer at the advancing menisci of the imbibition front, is reported for n-tetracosane confined in porous Vycor. For the rheology of a rod-like liquid nematogen (8OCB) we found no hints of the viscosity drop upon entering into the nematic phase, typical of the bulk rheology of this liquid crystal. 
\end{abstract}

\maketitle

\section{Introduction}
Spontaneous imbibition, that is the capillarity-driven invasion of a liquid in porous material is a very common phenomenon. From everyday life almost everybody knows the effect of coffee or tea being absorbed by a cube of sugar. A rather unpleasant instance of being confronted with its consequences is rising moisture in your basement walls. Another well-known example is the transport of water from the roots of a tree up to its limbs, branches and leaves, which is partly driven by capillary action \cite{Koch04}.

There has been significant progress with regard to a quantitative understanding of this phenomenon, most prominently with regard to the transport through rocks and soils,  starting with the seminal work of the French engineer Henry Darcy (1803-1858) \cite{Sahimi1993, Mecke2005}. Almost all experimental and theoretical studies, published to date, deal however with porous structures where the characteristic length scales, meaning the typical pore diameters, are on the macroscale, that is much larger than a typical size of the basic building blocks of the flowing liquids \cite{Alava04}.

After a short review of, and introduction into, the imbibition phenomenology of liquids in porous media in general, we will present a study on the spontaneous imbibition of liquids in pores a few nanometers across. We will vary the complexity of the building blocks of the liquids from water to simple short-chain n-alkanes and silicon oils to rod-like liquid crystals. Examinations on the phase transition behavior of nanopore-confined liquids will be part of this study as well. The goal is to scrutinize which principles regarding the rheology of these liquids, known from the bulk state, can be safely transferred to the transport across this nano- or mesoscopic geometries. %Moreover, we shall elucidate to what extent transport via the vapor phase, in the case of volatile liquids, competes with capillarity-driven transport in the pore center. 
%\rano{influence of vapor condensation?}

%Motivation
Given the emerging interest in micro- and nanofluidic applications \cite{Sparreboom09} such a study on the non-equilibrium behavior of liquids in extreme spatial confinement, on the nanometer scale, is not only of fundamental interest, but also of rather practical importance \cite{Eijkel05}. Moreover, the increasing use of nanoporous matrices as hard templates \cite{Thomas08} for the preparation of well-defined nanoscopic, soft-matter structures, such as nanotubes and nanorods \cite{Luo04}, necessitates a profound understanding of this phenomenology.
%\rano{Vielleicht doch besser nanoporous als mesoporous}

%The examples given above represent a small selection of the addressed topologies such as compressed sugar grains, cracks and pores in masonry blocks or concrete or the xylem network in trees. And finally these structures are also represented in the tortuous pore network in the porous \V glass, which is used in this thesis for examinations of the flow dynamics of liquids in spatial confinement.

%It is obvious that such modified boundary conditions and thereby altered flow rates play a crucial role in systems that are highly dominated by fluid-substrate interfaces. This applies for extremely miniaturized systems such as lab-on-a-chip applications, in which fluid amounts of some picoliters only are manipulated. The enormous academic and economic interests on the interfacial behavior of liquids are manifested by a vast publication rate concerning this issue during the last decade. Many different techniques like SFA, atomic force microscopy (AFM), particle image velocimetry (PIV), fluorescence recovery after photobleaching (FRAP) and controlled dewetting as well as molecular dynamics (MD) or lattice Boltzmann simulations were utilized.  \cite{gruener10a} 

\section{Fundamentals of liquid imbibition in nanoporous solids}    \label{sectionFundamentals}

\subsection[Liquid flow in isotropic pore networks: Darcy's law]{Liquid flow in isotropic pore networks: Darcy`s law}  \label{sectionDarcysLaw}
In order to introduce the fundamentals of liquid flow in nanoporous media, we start with a much simpler phenomenology: The flow of a liquid through a pipe is a prime example for the direct application of the Navier-Stokes equation resulting in the famous law of Hagen-Poiseuille. For a given pressure difference $\Delta p$\label{pressuredifference} applied along a cylindrical duct with radius $r$\label{radius1} and length $\ell$ the volume flow rate $\dot{V}$ is determined by
\begin{equation}
\dot{V} = \frac{\pi \,r^4 }{8\,\eta \,\ell} \,\Delta p \; .
\label{eq:HagenPoiseuille}
\end{equation}
Here $\eta$\label{viscosity} denotes the dynamic viscosity of the flowing liquid. In the following we will present a simple concept that allows one to extend the applicability of \rel{eq:HagenPoiseuille} to the flow through a complex pore network.

%In the next step one has to evolve concepts in order to account for the sponge-like structure of an isotropic pore network. %For this purpose a set of quantities is required that permits a sufficient characterization of such a matrix. %The most prominent approach is the concept of the tortuosity $\tau$, which will be introduced in the following.

\subsubsection{Characterization of a pore network}
Porous Vycor glass (code 7930) provided by Corning Incorporated was used over the course of this study. It is produced through metastable phase separation in an alkali-borosilicate system followed by an extraction of the alkali-rich phase. The final glass consists of a sponge-like network of tortuous and interconnected pores with mean pore radii on the nanometer scale embedded in a matrix mostly consisting of SiO$_2$ \cite{Elmer92}. 

In general such an isotropic pore network can be characterized by three quantities. The mean pore radius $\subs{r}{0}$\label{meanporeradius} and the volume porosity $\subs{\phi}{0}$\label{volumeporosity} are probably the most intuitive ones among them. In the course of this study two different batches of the monolithic glass were applied that differed in $\subs{r}{0}$ whereas they coincided in $\subs{\phi}{0} \approx 0.3$. For convenience we will refer to them as V5 ($\subs{r}{0} = 3.4$\,nm) and V10 ($\subs{r}{0} = 4.9$\,nm) from now on. These matrix properties were accurately ascertained employing nitrogen sorption isotherms conducted at 77\,K and through a subsequent ana\-ly\-sis within a mean field model for capillary condensation in nano- and mesopores. More details on the matrix characterization and preparation are given in Ref.~\cite{gruener10a}.

With only these two parameters a porous cuboid with edge length $a$ (and cross-sectional area $A=a^2$) consisting of\label{crosssectionalarea}
\begin{equation}
n=\frac{\subs{\phi}{0}\,A}{\pi\,\subs{r}{0}^2} \qquad \left(\;\Leftrightarrow \;\; \subs{\phi}{0} \equiv \frac{\subs{V}{void}}{\subs{V}{sample}} = \frac{n\,\pi\,\subs{r}{0}^2\,a}{a^3} \; \right)
\label{eq:porenumber}
\end{equation}
cylindrical pores with radius $\subs{r}{0}$ and length $a$ can be constructed. Assuming the capillaries to be aligned in flow direction the flow rate through the whole matrix is then given by $n$ times the single pore flow rate Eq.~(\ref{eq:HagenPoiseuille}) with $r=\subs{r}{0}$ and $\ell=a$. However, so far this description still lacks information on the orientation of the pores. 

\begin{figure}[!b]
\centering
\includegraphics*[width=.2\linewidth]{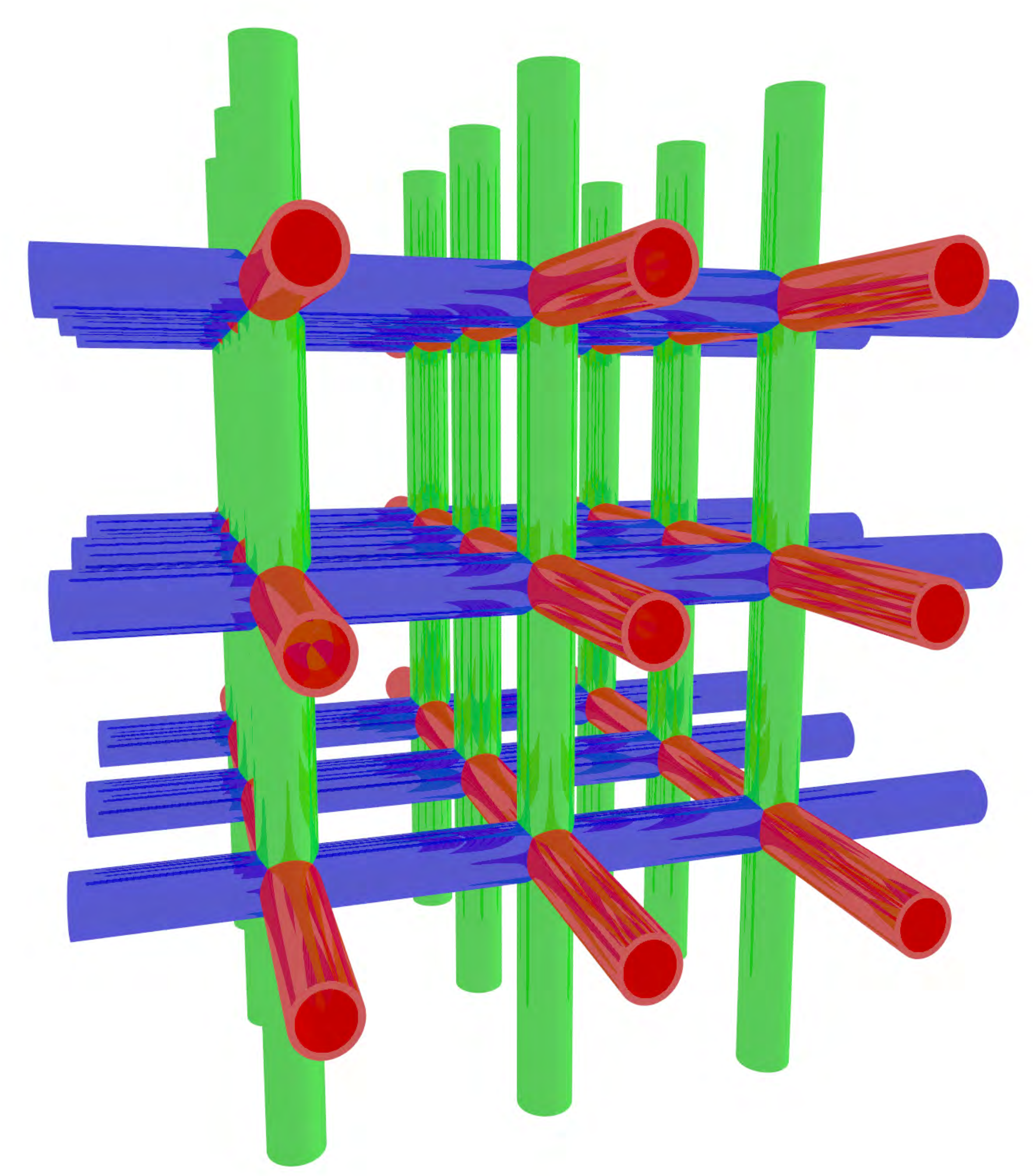}
\hspace{.1\linewidth}
\includegraphics*[width=.25\linewidth]{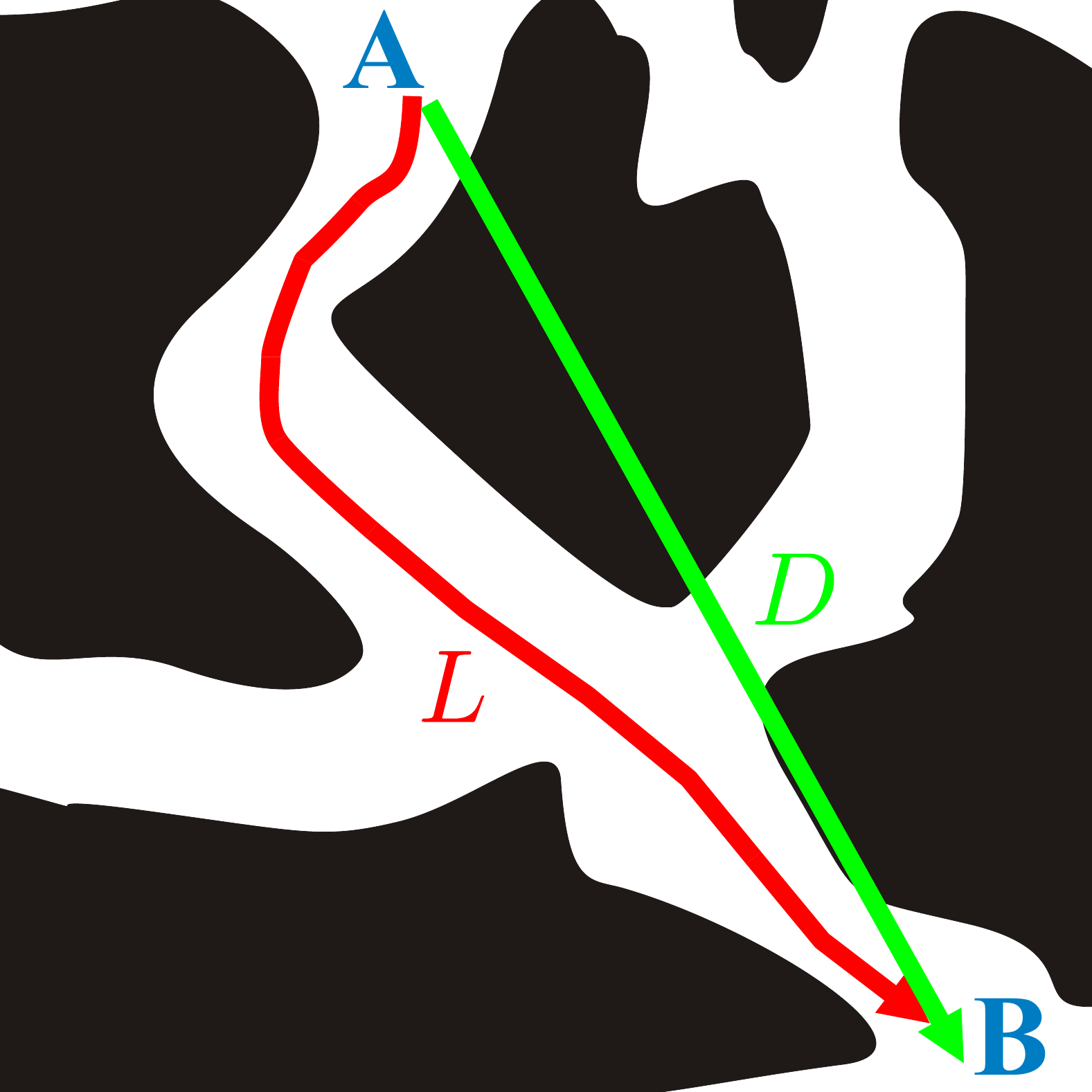}
\caption[Illustration of the meaning of the tortuosity of a pore network]{Illustration of the meaning of the tortuosity $\tau$ of a pore network with isotropically oriented (left) and meandering (right) pores.}
\label{pic_tortuosity}
\end{figure}

To account for the isotropy of the network as indicated in Fig.~\ref{pic_tortuosity}\,(left) it is necessary to introduce a third parameter, the so-called tortuosity $\tau$\label{tortuosity} along with the transformation
\begin{equation}
\dot{V} \quad \longrightarrow \quad \frac{1}{\tau} \dot{V}
\label{eq:tautransform}
\end{equation}
of the volume flow rate Eq.~(\ref{eq:HagenPoiseuille}). Pores totally aligned in flow direction would yield $\tau=1$, whereas isotropic distributed pores would result in $\tau=3$. This can be seen very easily. For a random orientation only every third pore is subjected to the pressure gradient and hence contributes to the flow. Therefore the net flow rate has to be divided by the factor three. 

To date serveral techniques have been applied to extract the tortuosity of the isotropic pore network in \V glass. Deducing the diffusion coefficient of hexane and decane by means of small angle neutron scattering (SANS) measurements $\tau$ was found to be in the range of 3.4 - 4.2 \cite{Lin92}. Gas permeation measurements performed with an in-house apparatus resulted in $\tau = 3.9\pm 0.4$ \cite{Bommer08}. Finally, calculations based on three-dimensional geometrical models yielded a value of approximately 3.5 \cite{Crossley91}. 

Interestingly, all values show a significant deviation from $\tau=3$ as derived from the previous considerations. Accordingly, there must be an additional aspect of the geometry that has so far been neglected. Regarding Fig.~\ref{pic_tortuosity}\,(right) this issue is apparent: the pores are not straight but rather meandering. In consequence the length $L$ of the path from any point A to another point B is always larger than the length $D$ of the direct interconnection of the two points. To correct the pore length for the larger flow path an additional factor of $\frac{L}{D}$ for the tortuosity must be introduced. Assuming $\tau =3.6$ this consideration yields for the \V pore network $L\approx 1.2\, D$. This result can vividly be interpreted as follows: the shortest way from the bottom of the previously introduced sample cuboid to its top is about 20\,\% longer than its edge length $a$.

\subsubsection{Darcy's law}

With all the preceding considerations in mind one is able to derive an expression that describes the flow of a liquid through a porous network. For a given porous matrix with cross-sectional area $A$ and thickness $d$\label{samplethickness} (along which the pressure drop $\Delta p$ is applied) the normalized volume flow rate $\frac{1}{A}\dot{V}$ is determined by
\begin{equation}
\frac{1}{A} \, \dot{V} = \frac{K}{\eta\,d} \,\Delta p \; .
\label{eq:Darcy}
\end{equation}
This expression is also known as Darcy's law \cite{Debye59}. The proportionality constant $K$ is the so-called hydraulic permeability\label{permeability} of the matrix. It is given by
\begin{equation}
K=\frac{\subs{\phi}{0}}{8\, \tau} \,\subs{r}{0}^2 \, .
\label{eq:permeability}
\end{equation}
At this point it must be emphasized that the permeability is solely specified by the matrix' internal structure and consequently it should be independent of the liquid and of the temperature. 

\subsection[Influence of the spatial confinement]{Influence of the spatial confinement}
So far we have completely neglected that the mean pore diameters of the pore network investigated here are orders of magnitude smaller than characteristic in usual flow paths in common miniaturized fluid manipulating applications. Indeed, the pore radii are merely 10 to 100 times larger than typical molecular diameters of simple liquids like water. Within the systematic study of chain-like hydrocarbons, which will be presented in section \ref{sectionHydrocarbonImb}, the dimensions of the liquid's building blocks even exceed the channel's diameter. For that reason it is evident that some questions about the influence of the confinement on the fluid dynamics arise. In the following the two most apparent ones will be discussed.

Though, beforehand, we will point out a remarkable feature being inherent in micro- and nanofluidic devices and applications. Because of the tiny characteristic length scales $\mathcal{L}$ the Reynolds number\label{charlengthscale1}
\begin{equation}
{\rm Re} \equiv \frac{\rm inertial\,\,\,force}{\rm viscous\,\,\,force}  = \frac{\rho\,v\,\mathcal{L}}{\eta}
\label{eq:Reynoldsnumber}
\end{equation}
(with the liquid's density $\rho$)\label{volumedensity} usually fulfills ${\rm Re} \ll 1$. Thus, flows in such restricted geometries are most likely laminar rather than turbulent. From this point of view the implementation of the Hagen-Poiseuille law seems to be justified. But, for very high flow speeds $v$\label{advancementspeed} this assumption does not remain valid. In section~\ref{sectionShortTime} it will be shown that this fact plays a crucial role in the very initial phase of a capillary rise experiment.

\subsubsection{Validity of continuum mechanical theory}

Up to now we have assumed the law of Hagen-Poiseuille to be valid even in pores with diameters below 10\,nm. However, one must not forget that this law is based on the principles of continuum mechanical theory, in which the behavior of a fluid is determined by collective properties such as the viscosity $\eta$ and the surface tension $\sigma$. This assumption certainly holds for ensembles of $10^{23}$ molecules. But within the pore confinement such amounts are not reached. This can easily be seen in the following example. Assuming water molecules to be spheres with a radius of 1.5\,\AA\ in a hexagonal close-packed structure one arrives at only $1000$ molecules per cross-sectional area. As a consequence, the validity of the continuum theory has to be put into question. 

On this score especially the development of the surface force apparatus (SFA) has stimulated extensive studies over the last three decades. The mobility of water and several hydrocarbons in extremely confined films was examined by experiment \cite{Israelachvili86, Horn89, Raviv01} and in theory \cite{Gupta97}. These studies revealed a remarkable robustness of the liquids' fluidity down to nanometer and even subnanometer spatial confinement. Moreover the validity of macroscopic capillarity conceptions at the nanoscale was demonstrated \cite{Fisher81, Fradin00}. The measurements within this study will provide further hints whether the concepts of viscosity and surface tension still remain valid in nanopore confinement.

\subsubsection{Validity of the no-slip boundary condition}  \label{subsectionNSBC}
The law of Hagen-Poiseuille implies the no-slip boundary condition. This means that the velocity of the fluid layers directly adjacent to the restricting walls equal the velocity of the walls themselves. Nowadays it is indisputable that this assumption does not hold unreservedly. Already 60 years ago Peter Debye and Robert Cleland introduced both slipping and sticking fluid layers at the pore walls in order to interpret a seminal experiment on liquid flow across porous \V \cite{Debye59}. In that way, they were able to quantitatively account both for increased as well as for decreased measured flow rates (compared to the predicted ones) within their examinations of the flow of hydrocarbons through porous \Vend.

\begin{figure}[!ht]
\centering
\includegraphics*[width=.5\linewidth]{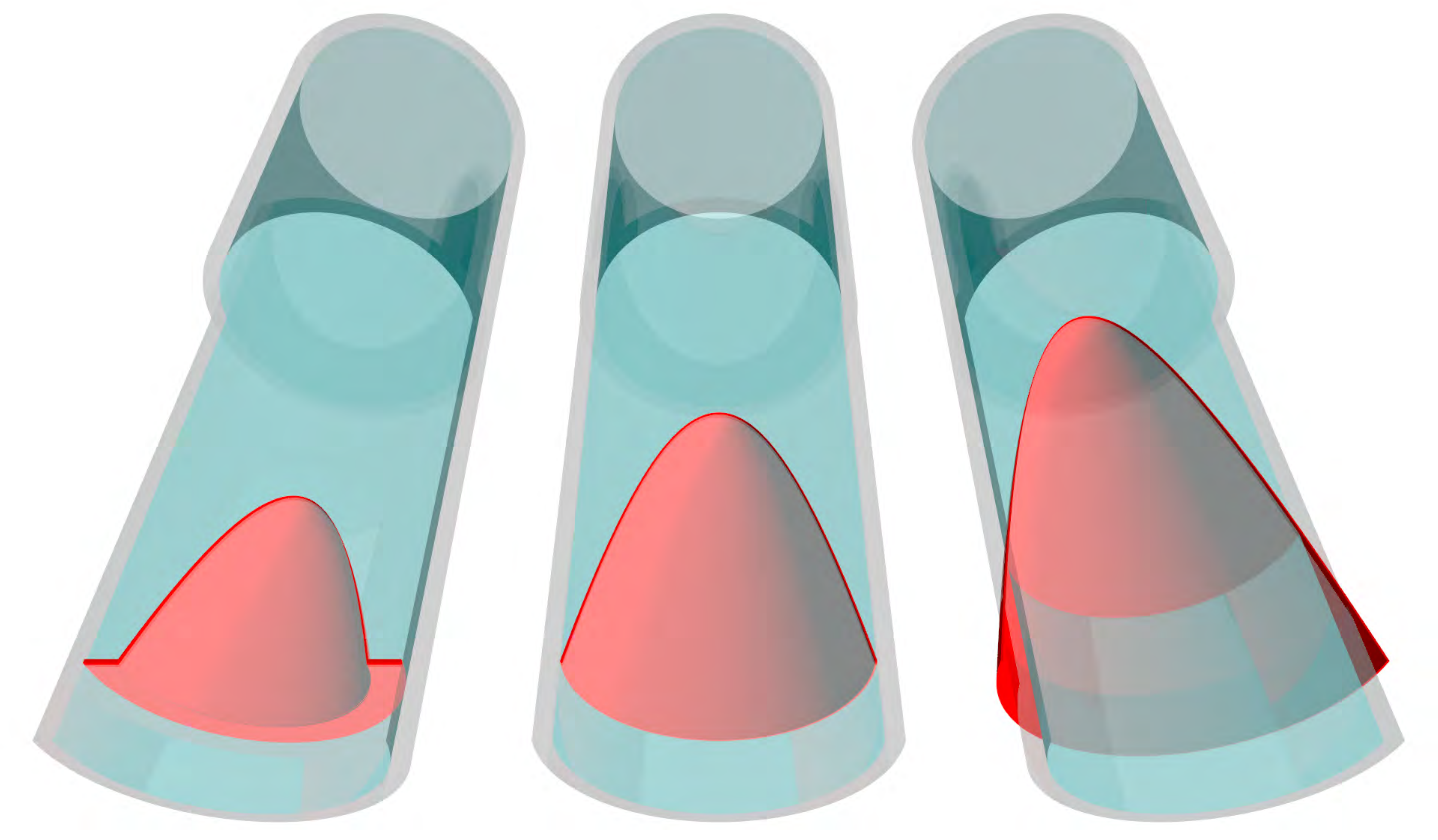}
\caption[Boundary conditions]{Illustration of the possible boundary conditions along with the corresponding parabolic velocity profiles in a cylindrical tube with radius $\subs{r}{0}$. Mass transport takes place only where the streaming velocity is different from zero. ({left}): The reduction of the net flow rate is due to sticking layers at the pore walls, which do not participate in the mass transport. In addition the maximum velocity in the pore center is smaller than for no-slip boundary conditions ({middle}) because of the smaller hydrodynamic pore radius $\subs{r}{h}<\subs{r}{0}$. This gives rise to a further dramatic decrease in the flow rate. ({right}): In contrast, a slipping liquid with a hydrodynamic pore radius $\subs{r}{h}>\subs{r}{0}$ causes the highest streaming velocity and consequently the highest net flow rate.}
\label{stickslip}
\end{figure} 
The concepts of a sticking and of a slipping liquid compared to the traditional no-slip boundary condition are exemplified in Fig.~\ref{stickslip} for a cylindrical tube with radius $\subs{r}{0}$. The degree of slip can be quantified by the slip length $b$\label{sliplength} with $\subs{r}{0} \equiv \subs{r}{h}-b$. The hydrodynamic pore radius $\subs{r}{h}$\label{hydrodynamicradius} measures the distance from the pore center to the radius where the streaming velocity reaches zero. In this representation the sticking layer boundary condition is indicated by a negative slip length $b$ whereas a positive slip length is typical of a slip boundary condition. The standard no-slip condition yields $b=0$ meaning $\subs{r}{0} = \subs{r}{h}$. 

Up to date many factors have been found that seem to influence the boundary conditions. The most prominent and maybe the least controversially discussed amongst them is the fluid-wall interaction expressed in terms of the wettability \cite{Barrat99, Pit00, Cieplak01, Tretheway02, Cho04, Schmatko05, Fetzer07, Voronov08, Maali08}. The weaker the interaction is the more likely is slip. In addition, shear rates beyond a critical value are supposed to induce slip, too \cite{Zhu01, Craig01, Priezjev04, Priezjev07}. In contrast, the influence of surface roughness is rather debatable \cite{Vinogradova06}. There are results for a decrease \cite{Zhu02, Pit00} as well as for an increase \cite{Bonaccurso03} of the slip length with increasing surface roughness. Furthermore, dissolved gases \cite{Granick03, Dammer06}, the shape of the fluid molecules \cite{Schmatko05} or the add-on of surfactants \cite{Cheikh03} seem to influence the boundary conditions. To sum up, there is a huge set of factors (see Refs.~\cite{Granick03, Lauga05, Neto05} for some good reviews) and certainly a complex interplay between many of them finally determines the interfacial behavior. 

The permeability $K$ of the membrane reflects such modified velocity boundary conditions:
\begin{equation}
K= \frac{\subs{\phi}{0}}{8\,\tau} \,\frac{\subs{r}{h}^4}{\subs{r}{0}^2} = \frac{\subs{\phi}{0}}{8\,\tau} \,\frac{(\subs{r}{0}+b)^4}{\subs{r}{0}^2} \; .
\label{eq:permeability2}
\end{equation}
Equation~\eqref{eq:permeability2} illustrates the high sensitivity of $K$ on $b$, provided $b$ is on the order of or even larger than $\subs{r}{0}$. Therefore, measuring the hydraulic permeability gives direct access to the slip length $b$ for a given liquid under given conditions. 

One has to keep in mind that boundary conditions and fluid properties derived from measured flow rates are subject to a central restriction: one cannot verify the predefined parabolic shape of the velocity profile in the nanoscopic flow geometry. This is because there is no direct access to the profile itself but only to flow rates, which correspond to the velocity profile integrated over the whole pore cross-sectional area. Nevertheless, molecular dynamics simulations prove the formation of parabolic flow profiles even down to channel radii of three molecular diameters \cite{Todd95, Travis97, Binder07} and, hence, justify inferences based on this major assumption.

\subsection{Laplace pressure \& capillary rise}
\begin{figure}[!ht]
\centering
\includegraphics*[width=.7\linewidth]{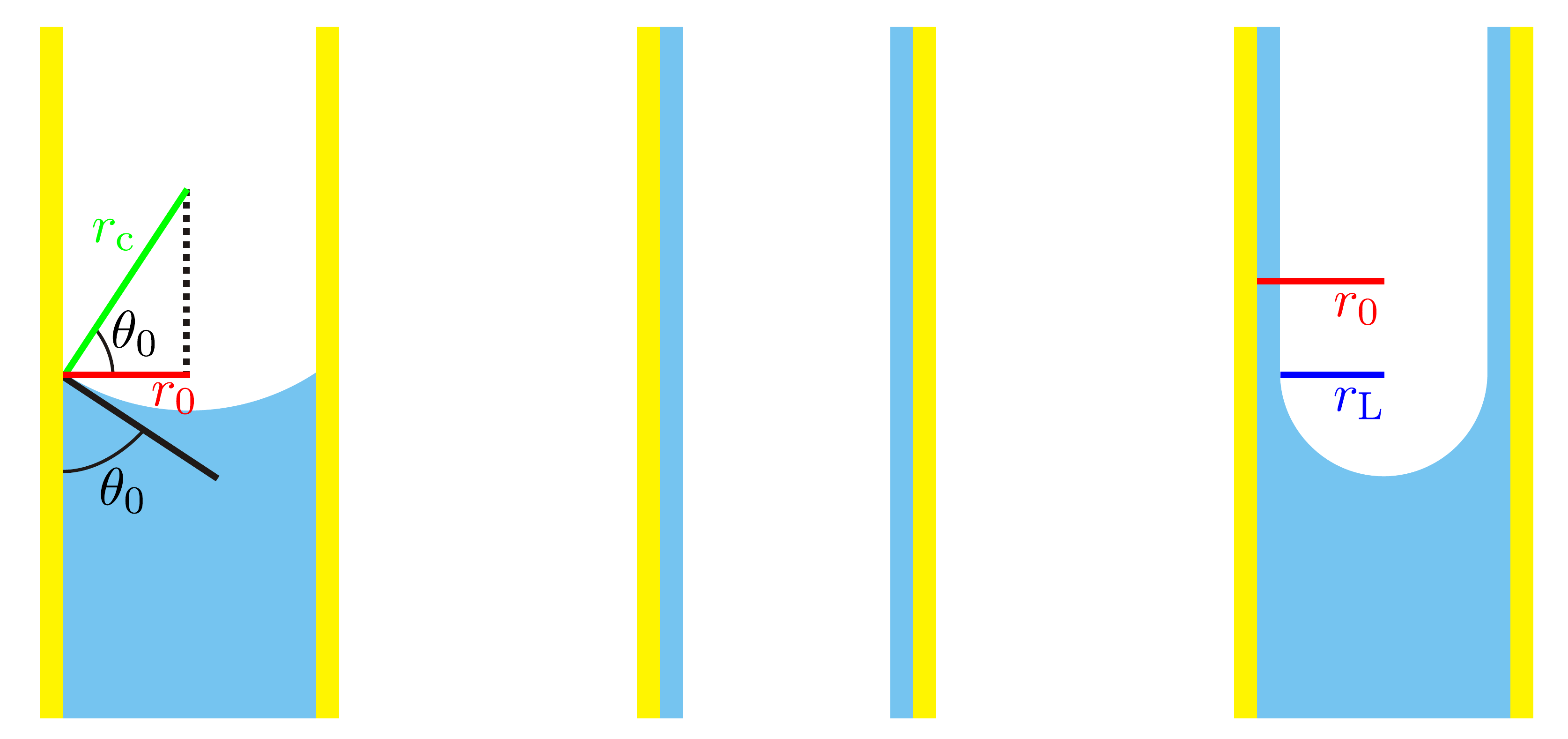}
\caption[Sketches of pores before and during an imbibition experiment]{Sketches of cuts through pores before and during an imbibition experiment. ({left}): Spherical meniscus of a liquid in a pore with the static contact angle $\subs{\theta}{0}$. From simple geometric considerations one can deduce the relation bet\-ween the radius of curvature $\subs{r}{c}$ and the pore radius $\subs{r}{0}$: $\subs{r}{c}=\frac{\subs{r}{0}}{\cos\subs{\theta}{0}}$. ({middle}): Water coating of the silica pore walls because of the finite humidity in the laboratory. ({right}): Illustration of the difference between the pore radius $\subs{r}{0}$ and the Laplace radius $\subs{r}{L}$ due to preadsorbed water layers, meaning that $\subs{r}{c}=\frac{\subs{r}{L}}{\cos\subs{\theta}{0}}$.}
\label{meniscus}
\end{figure}

From the physisist's point of view spontaneous imbibition is an impressive example for interfacial physics. The driving force behind the capillary rise process is the Laplace pressure $\subs{p}{L}$\label{Laplacepressure} acting on the curved meniscus of a liquid in a pore or porous structure. It is specified by
\begin{equation}
\subs{p}{L} = \sigma \left(\frac{1}{\subs{R}{1}} + \frac{1}{\subs{R}{2}} \right)
\label{eq:Laplace}
\end{equation}
with $\sigma$\label{surfacetension} being the surface tension of the liquid and $\subs{R}{1}$ and $\subs{R}{2}$ being the local radii of curvature of the liquid surface. For that reason it is important to obtain information on the shape of the meniscus. In general it is determined by an interplay of surface forces and gravity. The influence of the latter can be estimated by means of the so-called capillary length \label{capillarylength}
\begin{equation}
\subs{\lambda}{c} = \sqrt{\frac{\sigma}{\rho\,g}}
\label{eq:capillarylength}
\end{equation} 
with $g$ being the acceleration due to gravity and $\rho$ representing the density of the liquid. For water (with $\sigma\approx 72\,\frac{\rm mN}{\rm m}$ and $\rho=1\,\frac{\rm g}{{\rm cm}^3}$) Eq.~\eqref{eq:capillarylength} yields a capillary length of 2.7\,mm. For surfaces with characteristic lengths smaller than $\subs{\lambda}{c}$ the gravitational force can be neglected. Therefore, in the porous structure of \V with mean pore diameters on the order of 10\,nm the menisci are solely determined by surface forces. Consequently, they take the shape of a spherical cap (see Fig.~\ref{meniscus}) described by a unitary radius of curvature $R_1=R_2\equiv\subs{r}{c}$ resulting in a uniform Laplace pressure $\subs{p}{L}=\frac{2\,\sigma}{\subs{r}{c}}$ at each point of the meniscus. 

Eventually, from simple geometric considerations, one can deduce the relation bet\-ween the radius of curvature $\subs{r}{c}$ and the pore radius $\subs{r}{0}$ (see Fig.~\ref{meniscus}\,({left})) yielding 
\begin{equation}
\subs{p}{L} = \frac{2\,\sigma\,\cos\subs{\theta}{0}}{\subs{r}{0}}
\label{eq:Laplace2}
\end{equation}
for the Laplace pressure. It is obvious that the interaction between liquid and matrix expressed in terms of the static contact angle $\subs{\theta}{0}$ plays a crucial role in the capillary rise process. Spontaneous imbibition can only occur for a wetting or partially wetting liquid that is a liquid whose contact angle with the matrix is smaller than 90\degr. The influence of the contact angle hysteresis and a possible substitution of the dynamic contact angle $\subs{\theta}{D}$ for $\subs{\theta}{0}$ will be discussed in section~\ref{sectionShortTime}.

Of course, as the liquid rises beyond its bulk reservoir to a certain level $h$ the hydrostatic pressure $\subs{p}{h}= \rho\,g\,h$ acting on the liquid column increases. Accordingly the driving pressure $\Delta p$ must be modified to $\Delta p=\subs{p}{L}- \subs{p}{h}$. The final state is then derived from a balance between $\subs{p}{L}$ and $\subs{p}{h}$ \cite{Caupin08}. The maximum rise level $\subs{h}{max}$ is given by Jurin's law:
\begin{equation}
\subs{h}{max} = \frac{2\,\sigma\,\cos\subs{\theta}{0}}{\rho\,g\,\subs{r}{0}} \; .
\label{eq:hmax}
\end{equation}
It is worthwhile calculating $\subs{h}{max}$ for water in the silica network of \Vend. Assuming $\subs{\theta}{0} =0$\degr\, for water on a glass substrate and $\subs{r}{0}\approx 5\,$nm Eq.~\eqref{eq:hmax} yields $\subs{h}{max} \approx 2.9\,$km corresponding to a Laplace pressure of 290\,bar. Thus, for the here examined rise levels restricted by the maximum sample height to less than 5\,cm the gravitational force can be neglected meaning that 
\begin{equation}
\Delta p = \subs{p}{L} \;.
\label{eq:deltaP}
\end{equation}
The prevalence of surface forces over gravitation can likewise be expressed in terms of the dimensionless Bond number Bo:
\begin{equation}
{\rm Bo} \equiv \frac{\rm gravitational\,\,\,force}{\rm capillary\,\,\,force} = \frac{\rho\,g\,\mathcal{L}^2}{\sigma} \ll 1\; 
\label{eq:Bondnumber}
\end{equation}
because of the tiny characteristic length scales $\mathcal{L}$.

\subsection[Influence of preadsorbed liquid layers]{Influence of preadsorbed liquid layers}  \label{sectionPreadsLayers}

\begin{figure}[!t]
%\begin{minipage}[c]{0.3\textwidth}
\centering
\includegraphics*[width=.35\linewidth]{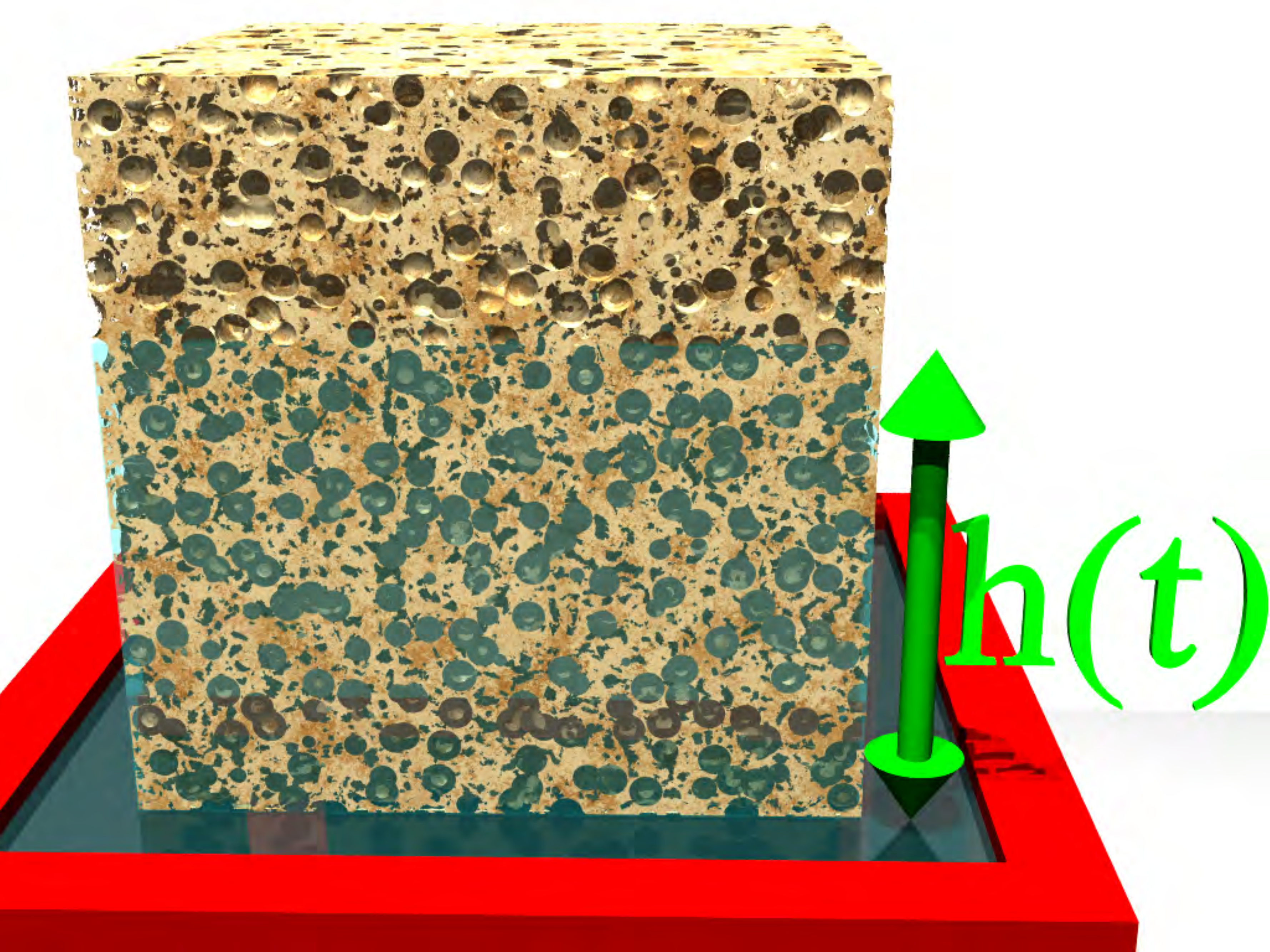}
%\end{minipage}
%\hspace{0.05\textwidth}
%\begin{minipage}[c]{0.3\textwidth}
%\centering
%\includegraphics*[width=.8\linewidth]{evaporation}
%\end{minipage}
\caption[\V sample during a capillary rise experiment]{Raytracing illustration of a \V sample during a capillary rise experiment filled up to the level $h(t)$.
%({left}): Raytracing illustration of a \V sample during a capillary rise experiment filled up to the level $h(t)$. ({right}): Illustration of the evaporation processes superimposing the imbibition measurement (black arrows). From both the liquid reservoir (blue arrows) and the already filled (green) parts of the sample (green arrows) liquid evaporates from the system (blue and green spheres) and might afterwards condense in the still empty (yellow) parts of the sample. Furthermore, liquid vapor from the advancing imbibition front might as well directly invade the empty parts beyond (red arrows and spheres).
}
\label{Vycor_filled}
\end{figure}

In order to gain information on the dynamics of the imbibition process we can now resort to Darcy's law (Eq.~\eqref{eq:Darcy}) discussed above. Considering Fig.~\ref{Vycor_filled}%\,({left})
one is able to conclude that the sample height $d$ appearing in Darcy's law has to be replaced by the actual rise level $h(t)$ since only the filled parts of the sample contribute to the flow dynamics. Moreover, at any given time $t$ the imbibed fluid volume $V(t)$ is closely related to $h(t)$ via
\begin{equation}
V(t)=\subs{\phi}{i} \,A\, h(t)\; .
\label{eq:LW1}
\end{equation}
Here $\subs{\phi}{i}$ denotes the initial volume porosity\label{iporosity} of the \V sample. This porosity is reduced with respect to $\subs{\phi}{0}$ because of water layers on the silica pore walls, which are immediately adsorbed under standard laboratory conditions because of the finite humidity and the highly attractive interaction between water and silica (see Fig.~\ref{meniscus}\,({middle})). This wall coating amounts to a 15\,\% to 20\,\% decrease in porosity. The adsorbed water is highly stabilized and can only be removed at elevated temperatures \cite{Wallacher97}.
% Especially the physisorbed first layer requires evacuation at temperatures $T>150$\dC\ for removal. 

The exact degree of coating depends on the sample temperature and the absolute humidity, thus it cannot be known beforehand. Fortunately $\subs{\phi}{i}$ can be extracted for each sample from the performed mass increase measurements, which will be presented later (see Fig.~\ref{imb_bsp}). Given the density $\rho$ of the imbibed liquid and the volume $\subs{V}{s}$\label{samplevolume} of the sample block one gains direct access to the initial porosity via the overall mass increase $M$\label{massuptake} due to the liquid uptake (indicated in Fig.~\ref{imb_bsp}):
\begin{equation}
\subs{\phi}{i} = \frac{ M}{\rho \, \subs{V}{s}} \;.
\label{eq:Phii}
\end{equation}

Similarly, the Laplace pressure must be influenced by the initial wall coating. The preadsorbed water layers necessarily lead to a reduction of the radius of curvature of the menisci (see Fig.~\ref{meniscus}\,({right})). This effect will be taken into consideration by substituting $\subs{r}{L}$\label{Laplaceradius} for $\subs{r}{0}$ in Eq.~\eqref{eq:Laplace2} with $\subs{r}{L}\le \subs{r}{0}$.
%, meaning
%\begin{equation}
%\Delta p = \subs{p}{L} =  \frac{2\,\sigma\,\cos\subs{\theta}{0}}{\subs{r}{L}}  \; .
%\label{eq:Laplace3}
%\end{equation}
We do not have any reliable information on the layer thickness $x=\subs{r}{0}-\subs{r}{L} $ and hence on the Laplace radius $\subs{r}{L}$. It can only be estimated from the difference between $\subs{\phi}{0}$ and $\subs{\phi}{i}$ assuming the pores to be perfect cylinders with a radius $\subs{r}{0}$. Then $\subs{r}{L}=\subs{r}{0}\, \sqrt{{\subs{\phi}{i}}/{\subs{\phi}{0}}}$ enables one to estimate the upper bound $x\le 5\,$\AA\ (depending on temperature and humidity). Based on this result we will assume $\subs{r}{L} = (\subs{r}{0} - 2.5\,$\AA$)\pm 2.5\,$\AA\ in order to account for this effect. Fortunately, the impact of $\subs{r}{L}$ on the overall imbibition dynamics is comparatively small (see \rel{eq:imb_gamma}) and, therefore, the uncertainty in the actual Laplace radius has only little effect on the final results.

\begin{figure}[!b]
\centering
\includegraphics*[width=.3\linewidth]{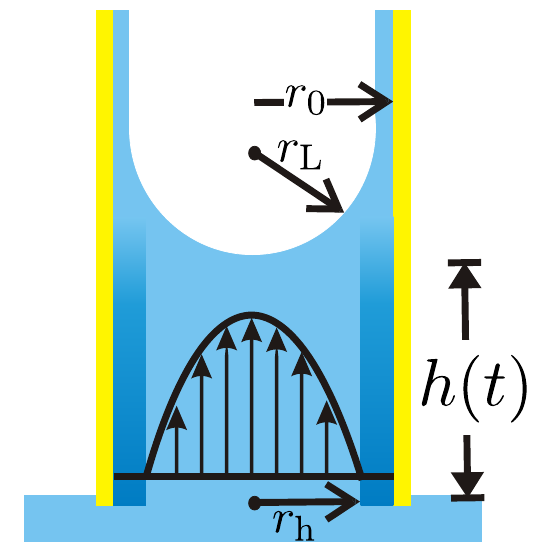}
\caption[Schematic view of the capillary rise in a cylindrical pore]{Schematic view of the capillary rise in a cylindrical pore with pore radius $\subs{r}{0}$, Laplace radius $\subs{r}{L}$ and hydrodynamic pore radius $\subs{r}{h}$.}
\label{imb_conf}
\end{figure}

\subsection[Dynamics of the imbibition process]{Dynamics of the imbibition process: BCLW law}
With all these preliminary considerations in mind we will now return to the dynamics of the imbibition process. Darcy's law Eq.~\eqref{eq:Darcy} in conjunction with Eq.~\eqref{eq:LW1} leads to the simple differential equation
\begin{equation}
\dot{h}(t)\, h(t) = \frac{K}{\subs{\phi}{i}\,\eta} \,\Delta p
\label{eq:LW2}
\end{equation}
solved through
\begin{equation}
h(t) = \sqrt{\frac{2\,K}{\subs{\phi}{i}\,\eta } \,\Delta p} \; \sqrt{t}\; .
\label{eq:LW}
\end{equation}
Nowadays this square root of time behavior is often referred to as the Lucas-Washburn (LW) law after its alleged discoverers \cite{Lucas18, Washburn21}. Actually, the $\sqrt{t}$-scaling was first found by Bell and Cameron \cite{Bell06} more than 10 years earlier \cite{Reyssat08}. Hence we will term Eq.~\eqref{eq:LW} BCLW law in order to acknowledge all contributors.

Given the driving pressure $\Delta p$ in accordance to Eq.~\eqref{eq:Laplace2} and the permeability $K$ corresponding to Eq.~\eqref{eq:permeability2} one obtains the rise level $h$ at a certain time $t$
\begin{equation}
h(t) = \underbrace{\sqrt{\frac{\sigma\,\cos\subs{\theta}{0}}{2\,\subs{\phi}{i} \, \eta}}\,\Gamma}_{\subs{C}{h}}\,\sqrt{t} \; .
\label{eq:imb_ht}
\end{equation}
Along with Eq.~\eqref{eq:LW1} and the density $\rho$ of the liquid Eq.~\eqref{eq:imb_ht} results in
\begin{equation}
m(t) = \underbrace{\rho\,A \,\sqrt{\frac{\subs{\phi}{i}\,\sigma\,\cos\subs{\theta}{0}}{2\,\eta}}\,\Gamma}_{\subs{C}{m}}\,\sqrt{t} \; ,
\label{eq:imb_mt}
\end{equation}
the mass increase $m$ of the sample due to the liquid uptake as a function of the time $t$. The corresponding flow configuration is illustrated in \bild{imb_conf}. We introduced the imbibition strength\label{imbability} $\Gamma$ in Eqs.~\eqref{eq:imb_ht} and \eqref{eq:imb_mt}, which is given through
\begin{equation}
\Gamma = \frac{\subs{r}{h}^2}{\subs{r}{0}} \, \sqrt{\frac{\subs{\phi}{0}}{\subs{r}{L}\,\tau}}  \; .
\label{eq:imb_gamma}
\end{equation}
Comparable to the hydraulic permeability it depends solely on the matrix' internal structure. %All liquid and temperature specific quantities and the sample's shape do not influence $\Gamma$, thus it should be a constant for all measurements with the same sample type. 

Interestingly, relation~\eqref{eq:imb_gamma} reveals that in principle the capillary rise dynamics are proportional to the square root of the capillary's radius: $\Gamma \propto \sqrt r$. As a consequence, in common capillary rise experiments with channel diameters on the order of some hundred micron the rise dynamics are so fast that without any additional instruments merely the steady-state equilibrium configuration at the very end of the process can be observed. This is contrary to imbibition in meso- and nanoporous networks. Here, the much smaller pores induce a lot more viscous drag, which reduces the overall dynamics significantly. Typical rise times in our experiments are on the order of some hours or even days. This fact allows for an easy recording of the dynamics by measuring the sample's mass increase $m(t)$ as a function of the time.

\subsection[Remarks on short time effects]{Remarks on short time effects}  \label{sectionShortTime}
Finally, we will outline some effects that dominate the imbibition process in its very initial phase, meaning the first few nanoseconds. However, these effects occur on time scales far beyond the temporal resolution limit of the experiment (on the order of one second) and therefore do not play a decisive role for the measurements. %But these phenomenons should not remain unmentioned. 

\subsubsection{BCLW Law in Nanopores}  
%To date the capillary rise behavior in meso- and nanopores has several times been studied by means of molecular dynamics simulations. At first glance the results are rather ambiguous. Some examinations prove the $\sqrt{t}$-law to be true \cite{Gelb02, Binder07} down to pores with radii less than 1\,nm. Nonetheless, a series of publications by Quirke \etal on imbibition in carbon nanotubes (CNT) suggests a linear function of the time \cite{Supple03, Supple04}. They attributed this behavior to the atomic smoothness of the nanotubes. But one has to keep in mind that in their simulated experiments it takes only 82\,ps until the tube is completely filled. Interestingly Binder \etal found a similar behavior restricted to a comparable short transient regime at the very beginning of the simulation followed by the classic BCLW law \cite{Binder07}. 
%
The BCLW law neglects inertial effects, thus it is only valid for time scales sufficient to establish viscous flow. For very short times a more general ansatz has to be applied (Bosanquet equation), which approaches the BCLW law for long times. The observed deviations from the BCLW law are possibly caused by a high degree of turbulences and initial deformations of the velocity profiles. This lasts until the flow reaches its typical low Reynolds numbers, for which turbulent flow cannot play a role at all. 

The typical time $\subs{\tau}{init}$ that is required for the viscous flow to establish itself in the pore can be estimated with \cite{deGennes04}
\begin{equation}
\subs{\tau}{init} \approx \frac{\rho\,\subs{r}{0}^2}{\eta} \; .
\label{eq:tauinit}
\end{equation}
This typically yields times on the order of some 10\,ps in high agreement with recent simulation studies \cite{Binder07, Supple03, Supple04, Gelb02}.

According to Ref.~\cite{Whitby07} the fluid-wall interaction, and in particular the surface friction due to molecular corrugation, additionally seem to influence $\subs{\tau}{init}$. For the atomic smooth CNT $\subs{\tau}{init}$ is rather increased and a lengthy non-BCLW behavior can be observed. The nearly frictionless surface of the nanotubes is also supposed to be responsible for the enhanced mobility of water and hydrocarbons flowing through these tiny ducts \cite{Majumder05, Holt06}. Nevertheless, this phenomenon occurs on time scales far beyond the temporal resolution limit of the experiment and does not affect our measurements at all \cite{deGennes04, Ralston08}. 

\subsubsection{Dynamic contact angle} \label{dynamiccontactangle}
The static contact angle $\subs{\theta}{0}$ of a drop resting on a surface is a constant determined by an interplay of the interactions between the solid, the liquid and the vapor phase. As a rule this assumption fails if the contact line begins to move. Here, a different contact angle, the so-called dynamic contact angle $\subs{\theta}{D}$ is observed. %It lies somewhere between the advancing (the largest achievable) and the receding (the smallest achievable) contact angle. Since $\theta$ can be increased or decreased (with respect to $\subs{\theta}{0}$) depending on whether the contact line begins to move in the direction of the gas phase or in the direction of the liquid phase this phenomenon is often referred to as contact angle hysteresis \cite{deGennes04, Sikalo05}.

It is this dynamic contact angle that is required as a boundary condition for modeling problems in capillary hydrodynamics. In consequence, it is important to estimate the changes that arise from this effect. Unlike the static contact angle, $\subs{\theta}{D}$ is not a material property. Actually, for small spreading velocities $v$ expressed in terms of the capillary number Ca
\begin{equation}
{\rm Ca} \equiv \frac{\rm viscous\,\,\,force}{\rm capillary\,\,\,force} = v\,\frac{\eta}{\sigma} \; ,
\label{eq:capillarynumber}
\end{equation}
it seems to solely be influenced by the capillary number itself \cite{Latva07}. This is the statement of the most widespread work relation describing the dynamic contact angle, namely the Hoffman-Voinov-Tanner law $\subs{\theta}{D}^3 \propto {\rm Ca}$. It is valid for ${\rm Ca} < 10^{-4} $ \cite{Ralston08} or, with some correction, for ${\rm Ca} < 10^{-2} $ \cite{Wang06}. 
%\begin{figure}[!t]
%\centering
%\includegraphics*[width=0.6\linewidth]{dca}
%\caption[Illustration of the dynamic contact angle]{Illustration of the dynamic contact angle. In the static state ({left}) on both sides of the drop the static contact angle $\subs{\theta}{0}$ is established. Though, on an inclined plane ({right}) the drop's shape becomes asymmetric with the dynamic contact angles $\subs{\theta}{r} <\subs{\theta}{0}$ and  $\subs{\theta}{a} >\subs{\theta}{0}$ on the receding and the advancing side, respectively.}
%\label{dca}
%\end{figure}
Interestingly, Richard Hofmann carried out a systematic study of dynamic contact angles in glass capillary tubes for a wide range of capillary numbers \cite{Hofmann75} and some years later Jiang \etal gave an empirical correlation for his results \cite{Jiang79}:
\begin{equation}
\frac{\cos \subs{\theta}{0}   -\cos \subs{\theta}{D}}{\cos \subs{\theta}{0} +1} = \tanh\left(4.96\cdot {\rm Ca}^{0.702}    \right) \; .
\label{eq:dynamiccontactangle}
\end{equation}

This expression can be used to estimate the impact of the contact angle hysteresis on the capillary rise experiments presented here. For this purpose one needs information on the prevailing capillary numbers being tantamount to the advancement speed $v$ of the liquid front. One can deduce this speed as the time derivative of Eq.~\eqref{eq:imb_ht}:
\begin{equation}
v (t) = \frac{{\rm d}h(t)}{{\rm d}t} = \frac{\subs{C}{h}}{2\,\sqrt{t}} \; .
\label{eq:risespeed}
\end{equation}
The obvious divergence for $t\to 0\,$s is again the manifestation of the deficiency of the BCLW law for very short time scales and has already been thematized in the literature \cite{Zhmud00, deGennes04}. 

%The flow speed $v(t)$ after one second elapsed rise time (consistent with the temporal resolution limit of the experiment) is a plausible quantity for an upper estimate of the prevailing capillary numbers. Furthermore, the highest speed will be observed for water (because of its relatively high surface tension) in V10 (because of the relatively low viscous drag due to its larger pore radii). Along with the respective measuring results (that will be presented in chapter~\ref{chapterRiseDynamics}) 
We obtain as the absolutely highest capillary number occuring in our experiments $\subs{{\rm Ca}}{max} \approx 10^{-6}$. Accordingly, the above approximation for low Ca Eq.~\eqref{eq:dynamiccontactangle} can be applied. With the static contact angle $\subs{\theta}{0}=0^\circ$ this finally yields the highest achievable dynamic contact angle $\subs{\theta}{D,max} = 2.2^\circ $ with $\cos \subs{\theta}{D,max} =0.9993 $ compared to $\cos \subs{\theta}{0} =1$. It is apparent that the phenomenon of the contact angle hysteresis needs no further consideration. This result agrees with MD studies of Martic \etal \cite{Martic02}. Indeed, they found an initial variation of $ \subs{\theta}{D}$ for liquids sucked into channels with 5\,nm radius. Nevertheless, it relaxed within some 10\,ns towards $\subs{\theta}{0}$ what is again far beyond the temporal resolution limit of the experiment.

What is more, the contact angle remaining $0^\circ$ during the whole measurement entails an interesting side effect. In a recent lattice Boltzmann simulation Ku\-su\-maat\-maja \etal \cite{Pooley08} examined the influence of surface patterning on the capillary filling of microchannels. They found that for $\theta <30^\circ$ the structuring has only little effect on the filling process and for completely wetting liquids there is no observable influence at all. Therefore we do not have to worry about atomic irregularities of the pore shape.

\section{Experimental}
%In this chapter the measuring techniques that were applied in order to study the imbibition dynamics of liquids in porous \V will be introduced. Both the experimental setups and the sample preparation for the respective method will be elucidated.

\begin{figure}[!t]
\centering
\includegraphics*[width=.59\linewidth]{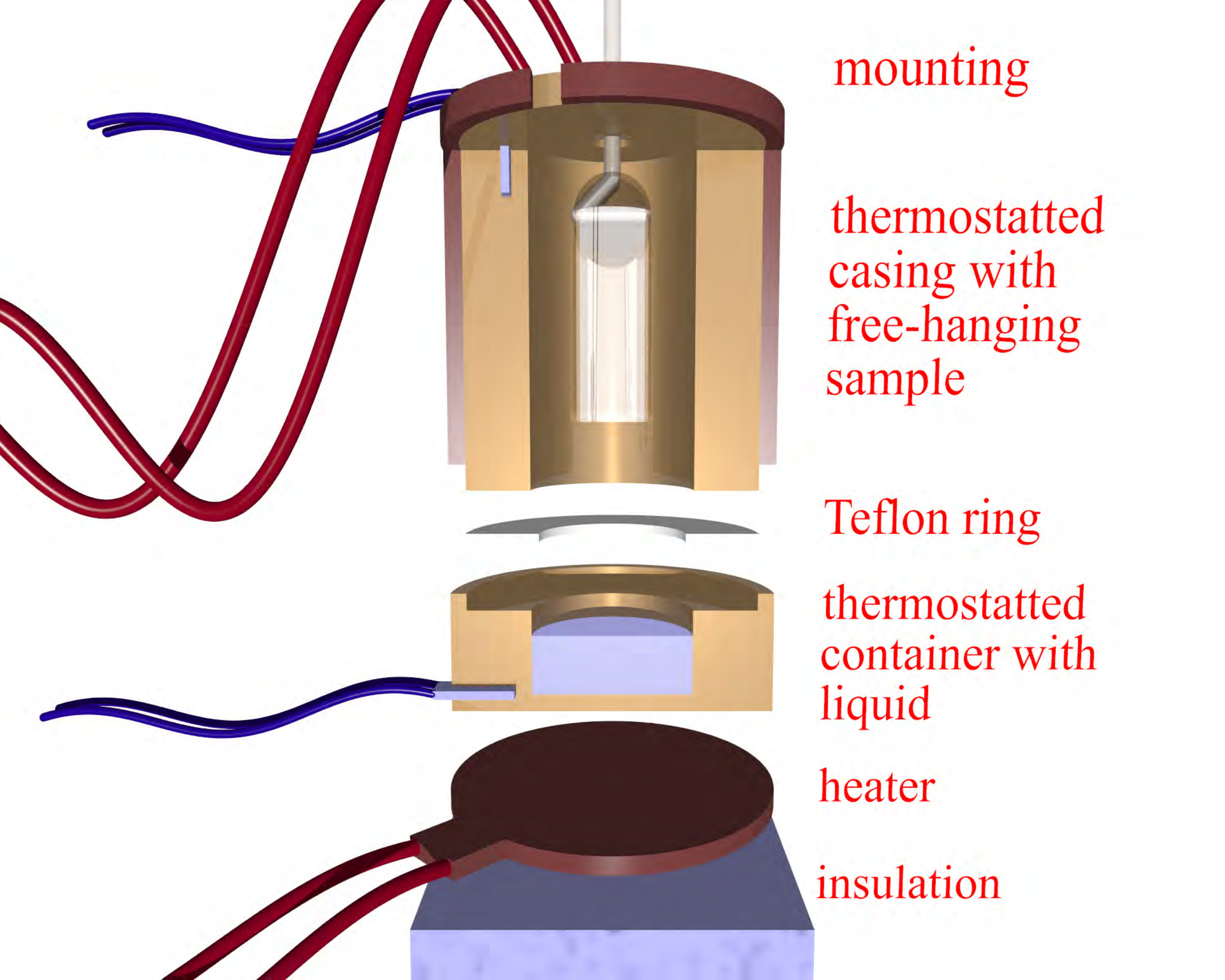}
\caption[Exploded view of the imbibition setup (LIS)]{Exploded view of the imbibition setup (raytracing illustration). The top of the \V sample is glued to a wire and hangs freely into the cell. The cell itself consists of a casing and a container. Both are build out of copper and can separately be thermostatted.}
\label{LIS_cell}
\end{figure}

Imbibition dynamics are studied via a recording of the samples' mass increase $m(t)$. The corresponding setup is depicted in Fig.~\ref{LIS_cell}. For a time-dependent measurement of the force acting on and, hence, of the mass increase of the porous \V block the sample is installed on standard laboratory scales applying a special mounting. 
In order to perform measurements beyond room temperature a cell was constructed that allows for a simultaneous thermostatting of the sample itself and the liquid reservoir beneath. 

The samples were cut into regular shapes of either cylinders (V5) or cuboids (V10) of known dimensions (cross-sectional area $A$ and height $\subs{h}{0}$). Prior to measurement they were stored in a desiccator to avoid the uptake of impurities from the surrounding air. 
\begin{figure}[!b]
\centering
\includegraphics*[width=0.8\linewidth]{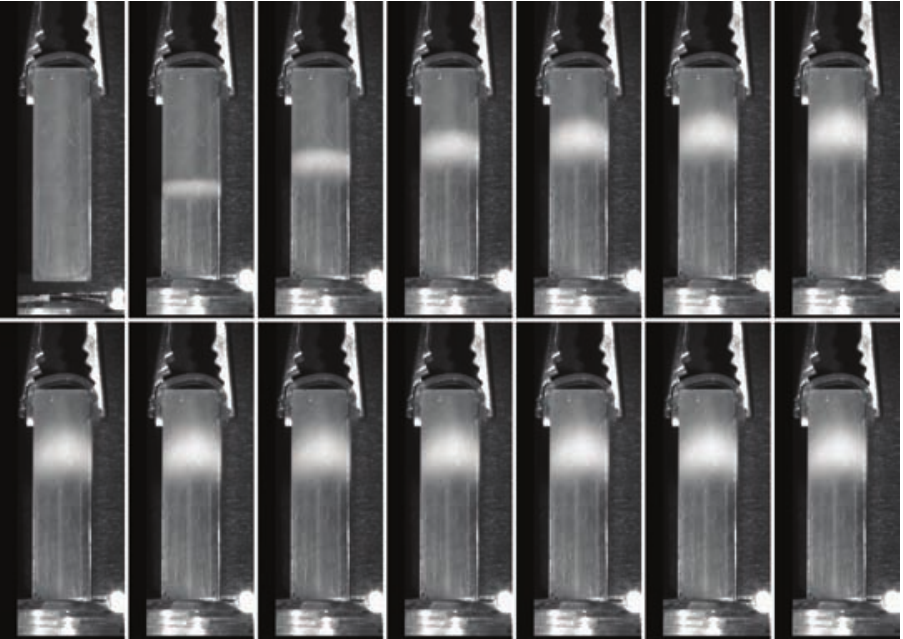}
\caption[Pictures of a pinning experiment of water invading V10]{Series of pictures of a pinning experiment of water invading a porous \V block (V10, $\subs{V}{s}=8.14\times 6.05\times 29.00\,$mm$^3$) at room temperature. The sample is not sealed. Intervals between pictures: 106\,min. Due to the increasing impact of evaporation the front is eventually pinned.}
\label{pix_pinning}
\end{figure}

For highly volatile liquids such as water and the n-alkanes up to decane a sealing of the samples' top and side facets is of utmost importance. Partly this is because of the possible influx of molecules via the vapor phase. However, with increasing rise level another problem gains importance, namely the evaporation from the sample. This would actually lead to a pinning of the rise level $h$ determined by a balance between the liquid supplying imbibition rate and the liquid detracting evaporation rate (see Fig.~\ref{pix_pinning}). For water such a sealing can very well be realized with Scotch tape. For the short-length n-alkanes a two-component adhesive was applied. For all other liquids the samples' facets remained unsealed.

\begin{figure}[!t]
\centering
\includegraphics*[width=.8\linewidth]{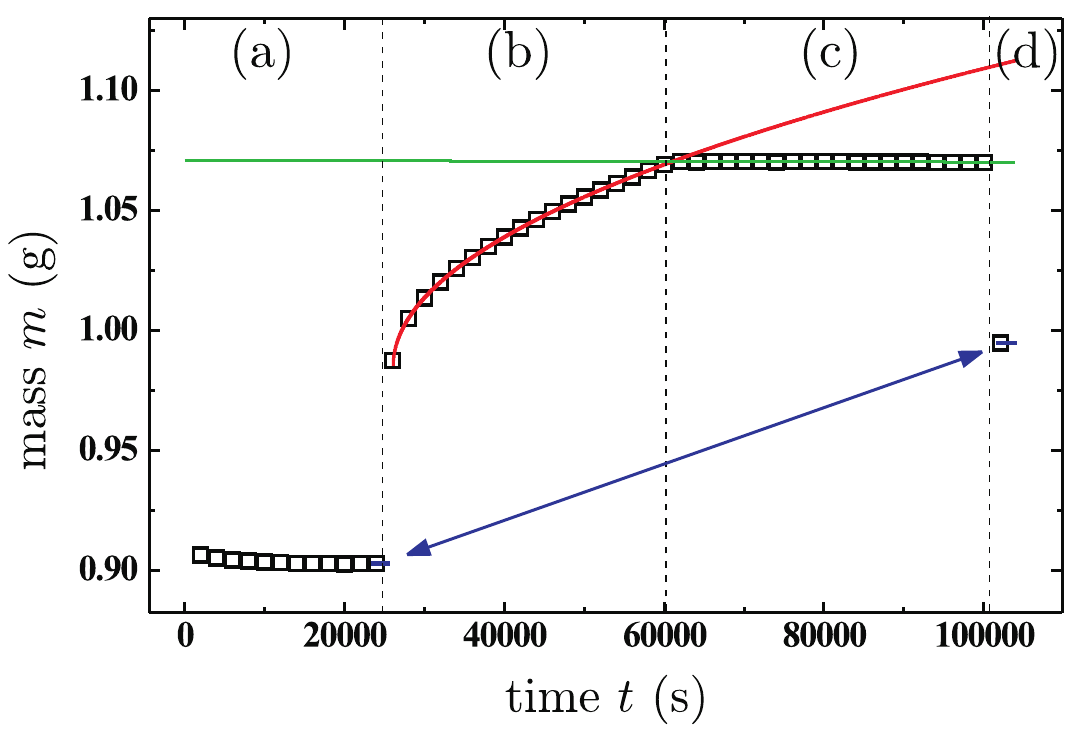}
\caption[Representative measurement of the sample's mass increase due to liquid imbibition]{Representative measurement of the mass increase of a porous \V block (V10, $\subs{V}{s}=6.25\times 5.05\times 12.70\,$mm$^3$) due to the imbibition of n-tetracosane (n-{C24H50}) at $T=59\,$\degr C. Solid lines are a $\sqrt{t}$-fit to the liquid's imbibition behavior and a linear fit to the level of saturation. The arrow indicates the difference between the sample's mass before ($\subs{m}{s}$) and after ($\subs{m}{e}$) the experiment. The experiment is divided into four regimes as indicated by the dashed lines and marked by (a) - (d). The data density is reduced by a factor of 2000.}
\label{imb_bsp}
\end{figure}

The measuring principle of the gravimetric experiments can nicely be illustrated referring to the representative mass increase measurement depicted in Fig.~\ref{imb_bsp}. Four distinct regimes are indicated. In the beginning the sample is put into the thermostatting cell and is mounted on the laboratory scale so that it hangs freely above the bulk reservoir. Regime (a) is now described by a mass decrease of the sample, which can be seen as the obvious manifestation of its thermalization. Due to the elevated
temperature (here $T=59\,$\degr C) the sample loses a part of its preadsorbed water and accordingly gets lighter. A sufficient thermalization is then given for an adequate convergence of the mass towards $\subs{m}{s}$. 

The measurement is started by moving the cell upward until the sample touches the liquid surface. Immediately a meniscus forms along the block's perimeter. This induces a force $\subs{F}{S}$ towards the reservoir acting on the sample, which becomes noticeable by a sudden jump in mass right at the beginning of regime (b). The connection between $\subs{F}{S}$ and the perimeter length $P$ is determined by
\begin{equation}
\subs{F}{S} = P\,\sigma\,\cos{\subs{\theta}{0}} 
\label{eq:FS}
\end{equation}
widely known from the Wilhelmy plate assembly for measuring the surface tension of a liquid. Here it enables one to quantitatively analyze the mass jump. Assuming $\subs{\theta}{0}=0$\degr\, and given the surface tension $\sigma \approx 30\,\frac{\rm mN}{\rm m}$ of n-tetracosane at $T=59$\,\degr C Eq.~\eqref{eq:FS} gives $\subs{F}{S}\approx 0.7\,$mN or, equivalently, an apparent mass jump of $\Delta m \approx 0.07\,$g. This is in high accordance with the measurement. %Yet, this jump can also be less distinctive since the counteracting buoyancy force sensitively depends on the actual depth of immersion. 

This jump is only an initial effect and can be seen as a constant offset that should not disturb the subsequent imbibition process itself. The latter is the outstanding effect in regime (b). It can be described by a $\sqrt{t}$-fit in accordance to Eq.~\eqref{eq:imb_mt}. Thus, the $\sqrt{t}$-fit provides $\subs{C}{m}$ and then Eq.~\eqref{eq:imb_mt} results in the imbibition strength 
\begin{equation}
\Gamma = \frac{\subs{C}{m}}{\rho\,A} \,\sqrt{\frac{2\,\eta}{\subs{\phi}{i}\,\sigma\,\cos\subs{\theta}{0}}} 
\label{eq:GammaM}
\end{equation}
where the initial porosity $\subs{\phi}{i}$ according to Eq.~\eqref{eq:Phii} and the bulk fluid parameters are applied. The contact angle is assumed to be 0$^\circ$ and consequently $\cos\subs{\theta}{0}=1$. The latter assumption was also tested experimentally and proved true for all liquids taken into consideration for this study by means of contact angle measurements.

At some point a level of saturation is entered. The deviation from the $\sqrt{t}$-law in regime (c) signals that the sample is completely filled. In regime (d) the sample is detached from the bulk liquid reservoir in order to extract the mass $\subs{m}{e}$ of the completely filled sample. The overall mass uptake is then $M=\subs{m}{e}-\subs{m}{s}$. 

%In addition to the before-mentioned method one can apply the rise level relation Eq.~\eqref{eq:imb_ht} in order to gain access to the imbibition strength of the sample. It can be deduced from the time $\subs{t}{0}$\label{fillingtime} that it takes for the complete filling of a sample with height $\subs{h}{0}$\label{sampleheight}. Eventually, one arrives at:
%\begin{equation}
%\Gamma = \frac{\subs{h}{0}}{\sqrt{\subs{t}{0}}} \,\sqrt{\frac{2\,\subs{\phi}{i}\,\eta}{\sigma\,\cos\subs{\theta}{0}}}  \equiv  \subs{\Gamma}{t(ime)}\; .
%\label{eq:GammaH}
%\end{equation}
%Though, it is important to keep in mind that $\subs{\Gamma}{t}$ is not derived from the rise level dynamics, but from one single known point: $h(\subs{t}{0}) = \subs{h}{0}$. The imbibition strength $\subs{\Gamma}{m}$ according to \rel{eq:GammaM} should therefore imply the more reliable results.% Nonetheless, in some special cases \rel{eq:GammaH} will be rather useful.

\section{Spontaneous imbibition of water in porous \Vend}

\begin{table}[!b]
\centering
\setlength\extrarowheight{2pt}
\caption[Imbibition strengths of water in porous \V deduced from gravimetric measurements]{Characterization of the imbibition dynamics of water in porous \V at three different temperatures by means of the imbibition strength $\Gamma$ as deduced from gravimetric measurements according to \rel{eq:GammaM} (in units of $10^{-7}\,\sqrt{\rm m}$).} 
\begin{tabularx}{1\textwidth}{>{\centering\arraybackslash}X>{\centering\arraybackslash}X>{\centering\arraybackslash}X} \toprule
 temperature & V5 & V10   \\  \midrule\midrule
  25\,$^\circ$C &   $126.7\pm 8.2$   &  $170.0\pm 7.9$  \\
  40\,$^\circ$C &   $120.5\pm 7.4$   &  $180.9\pm 11.6$  \\
  60\,$^\circ$C &   $125.3\pm 11.2$  &  $174.3\pm 11.4$ \\ \bottomrule
\end{tabularx}
\label{tab:Gamma_water}
\end{table}

Experiments on the capillary rise of water in V5 and V10 were conducted at three different temperatures $T$ (25\dC, 40\dC, and 60\dC). The results are summarized in \tab{tab:Gamma_water}. As predicted for the imbibition strength the obtained values are independent of the temperature $T$. The arithmetic averages over all temperatures are $(123.7\pm 2.1)\cdot 10^{-7}\,\sqrt{\rm m}$ for V5 and $(173.7\pm 3.1)\cdot 10^{-7}\,\sqrt{\rm m}$ for V10. According to \rel{eq:imb_gamma} these measured quantities correspond to hydrodynamic pore radii $\subs{r}{h} = (2.85\pm 0.21)$\,nm for V5 and $\subs{r}{h} = ( 4.41 \pm  0.28 )$\,nm for V10, or, in terms of the slip length $b = \subs{r}{h}-\subs{r}{0}$,

\begin{eqnarray}
b & \approx & -5 \,{\AA} \qquad  \mathrm{for\;\, water\;\, invading\;\, V5\;\, and}  \nonumber \\   
b & \approx & -5 \,{\AA} \qquad  \mathrm{for\;\, water\;\, invading\;\, V10.} \nonumber
\label{eq:sliplengthWater}
\end{eqnarray}
Consistently for both V5 and V10 a {\it negative} slip length of about 5\,\AA\ was found. Considering the diameter of a water molecule of approximately 2.5\,\AA\ this result can be interpreted as follows: two layers of water directly adjacent to the pore walls are immobile meaning that they are pinned and do not take part in the flow. Taking this into account (through the above-mentioned negative slip length) the imbibition dynamics can conclusively be described by the model proposed in section~\ref{sectionFundamentals}. In consequence one may conclude that the residual inner compartment of the liquid obeys classical hydrodynamics based on continuum mechanical theory. As a result, our experiments confirm former findings on the conserved fluidity \cite{Israelachvili86, Horn89, Raviv01} and capillarity \cite{Fisher81, Fradin00} of confined water -- except for the pinned layers.

The assumption of the compartmentation of nanopore-confined water just stated is corroborated by recent molecular dynamics studies on the glassy structure of water boundary layers in \V and the expected existence of sticky boundary layers in Hagen-Poiseuille nanochannel flows for strong fluid-wall interactions \cite{Heinbuch89, Ricci00, Gallo00, Vichit00, Gallo01, Castrillon09}. By means of X-ray diffraction distortions of the hydrogen-bonded network of water near silica surfaces were found \cite{Fouzri02}, which might be responsible for the markedly altered liquid properties. Tip-surface measurements document a sudden increase in the viscosity by orders of magnitude in 0.5\,nm proximity to hydrophilic glass surfaces \cite{Li07}. It also extends former experimental results with respect to the validity of the no-slip boundary condition for water/silica interfaces, in which this condition was proven down to at least 10\,nm from the surface \cite{Lasne08} whereas slip-flow for water is only expected at hydrophobic surfaces \cite{Vinogradova99, Lasne08}. 

Here, it is important to highlight that a significant influence of silica dissolution in water can definitely be ruled out. The resultant aqueous electrolyte (often referred to as polywater) would of course be characterized by altered liquid properties and could therefore likewise be responsible for the observed decreased overall invasion dynamics. Nevertheless, as was pointed out in detail by Tombari \etal \cite{Tombari05}, the dissolution of silica in water for the measuring times and temperatures considered here is vanishingly small and may be neglected.

Furthermore, any by-passing material transport via the vapor phase could be excluded. According to recently conducted lattice Boltzmann simulations such condensation processes would entail a breakdown of the usual imbibition dynamics \cite{Pooley09}. However, a significant influence of evaporation from and subsequent condensation beyond the advancing liquid front can be ruled out referring to thermodynamic argumentations \cite{Gruener09}. Moreover, recently performed neutron radiography measurements did not only verify this prediction, but also allow us to conclude that the imbibition dynamics are only marginally affected by this transport mechanism \cite{gruener10c}.

%The presented imbibition experiments entail additional rheologic details. As mentioned earlier Eqs.~\eqref{eq:imb_ht} and \eqref{eq:imb_mt} intrinsically assume a parabolic velocity profile across the pore cross section, which implies a linear variation in the viscous shear rate $\dot{\gamma}$, starting with 0\,$\frac{1}{\rm s}$ in the pore center to a $t$-dependent maximum at $\subs{r}{h}$: $\subs{\dot{\gamma}}{max} \propto \frac{1}{\sqrt{t}}$. For the V5-experiment, for example, one may estimate $\subs{\dot{\gamma}}{max}$ to decrease from $7\cdot 10^4\,\frac{1}{\rm s}$ after 1\,s to $3\cdot 10^2\,\frac{1}{\rm s}$ at the end of the experiment. Since there are no $t$-dependent, and consequently no $\dot{\gamma}$-dependent, deviations of $m(t)$ from a single $\sqrt{t}$-fit the measurements testify to the absence of any non-Newtonian behavior of water. Despite the relatively large $\dot{\gamma}$'s ascertained here, the latter is not too surprising. The viscous forces of $\mathcal{O} (\eta \,d^2\,\dot{\gamma})$ (with the molecular dimension $d$ on the order of a few \AA) can only overcome the strong water/silica interactions of $\mathcal{O} (A/d)$ (with the Hamaker constant $A \sim 10^{-19}  $\,J) for $\dot{\gamma} > 10^{12}\,\frac{1}{\rm s}$ \cite{Lauga05, Israelachvili06} -- significantly beyond the shear rates probed here.

Finally, it is worthwhile noting that water encounters a negative pressure, which linearly decreases from $-\subs{p}{L}$ (on the order of some hundred bar) at the advancing menisci to atmospheric pressure at the bulk reservoir. Water's hydrogen bridge bond network is expected to be responsible for an increase in viscosity $\eta$ and decrease in density $\rho$ under such large tensile pressures. Based on thermodynamic models for stretched water \cite{Stanley02, Tanaka00a}, one may estimate a $T$-dependent $\sim 3\,\%$ decrease in the overall dynamics due to this effect, which is, unfortunately, below the error margins. Finally, we would refer the interested reader to two more detailed publications reporting on capillary rise dynamics of water in nanoporous silica \cite{Gruener09, Huber07}.

\section[Spontaneous imbibition of hydrocarbons in porous Vycor]{Spontaneous imbibition of hydrocarbons in porous \Vend}  \label{sectionHydrocarbonImb}
\begin{figure}[!b]
\centering
\includegraphics*[width=.5\linewidth]{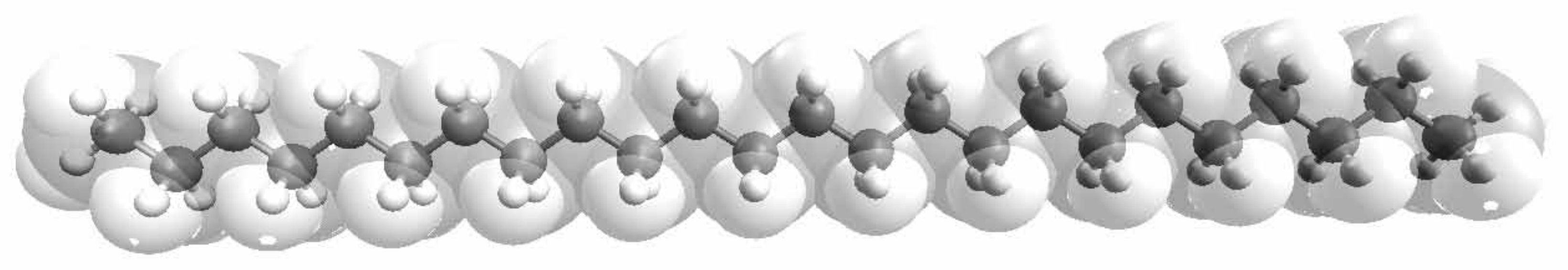}
\includegraphics*[width=.5 \linewidth]{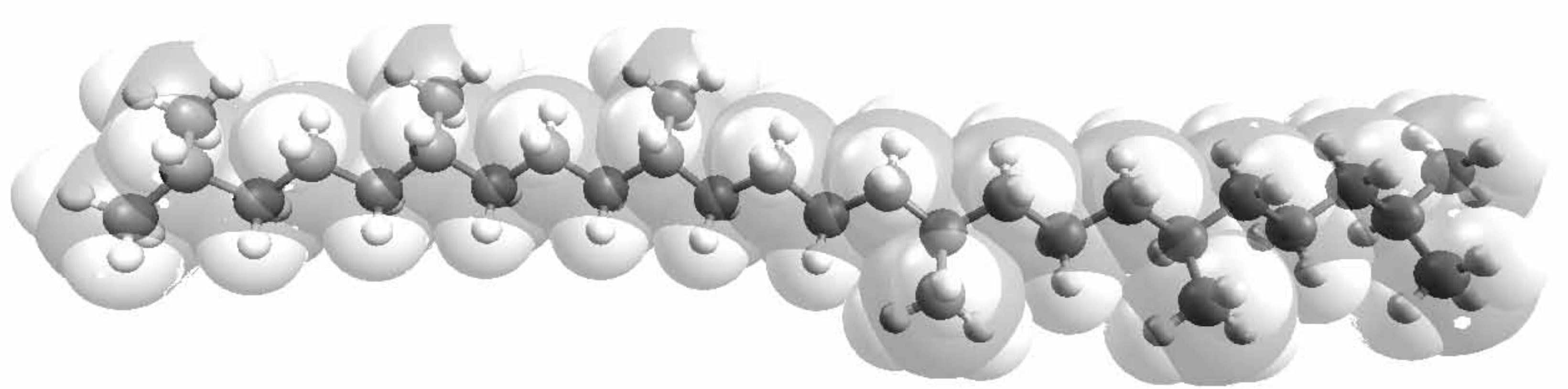}
\caption[Illustrations of hydrocarbons in all-trans configuration]{Illustrations of the all-trans configurations of the linear hydrocarbon n-tetracosane (n-C$_{24}$H$_{50}$) and the branched hydrocarbon squalane (C$_{30}$H$_{62}$). The backbones of both chains carry 24 C atoms, so they coincide in the all-trans molecule length $\ell\approx 3$\,nm.}
\label{alkanemolecules}
\end{figure} 
The imbibition study of water in porous \V glass just presented reveals many aspects of the behavior of a liquid in interface-determined systems. Anyhow, these results could just as well have been obtained on a planar glass substrate, meaning that so far no effects explicitly due to the liquids' restrictions to a cylindrical pore geometry have been reported. One might wonder whether the water molecules are too small as compared with the channel diameter to be subject to any limitations. This consideration motivates the hereinafter presented systematic study of the imbibition dynamics of chain-like hydrocarbons namely the homologous series of n-alkanes (n-C$_n$H$_{2n+2}$) and the branched hydrocarbon squalane.

The alkane molecule's all-trans length $\ell$\label{alltranslength} is proportional to the number of C atoms $n$ in the chain's backbone
\begin{equation}
\ell(n) = 2.1\,\text{\AA}+(n-1)\cdot1.25\,\text{\AA} \;
\label{eq:alkanelength}
\end{equation}
whereas their diameter is independent of $n$ (approximately 3\,\AA). Hence, the alkane series allows one to easily vary the aspect ratio of the molecule, while many important liquid properties (density $\rho$, surface tension $\sigma$, viscosity $\eta$) nearly remain the same \cite{Small86}. In particular all alkanes totally wet silica. For that reason, alkanes are particularly well suitable for a systematic study of the influence of the shape of the liquids' building blocks on the flow dynamics through channels with diameters comparable to the molecule's length.

\begin{figure}[!b]
\centering
\includegraphics*[width=.7\linewidth]{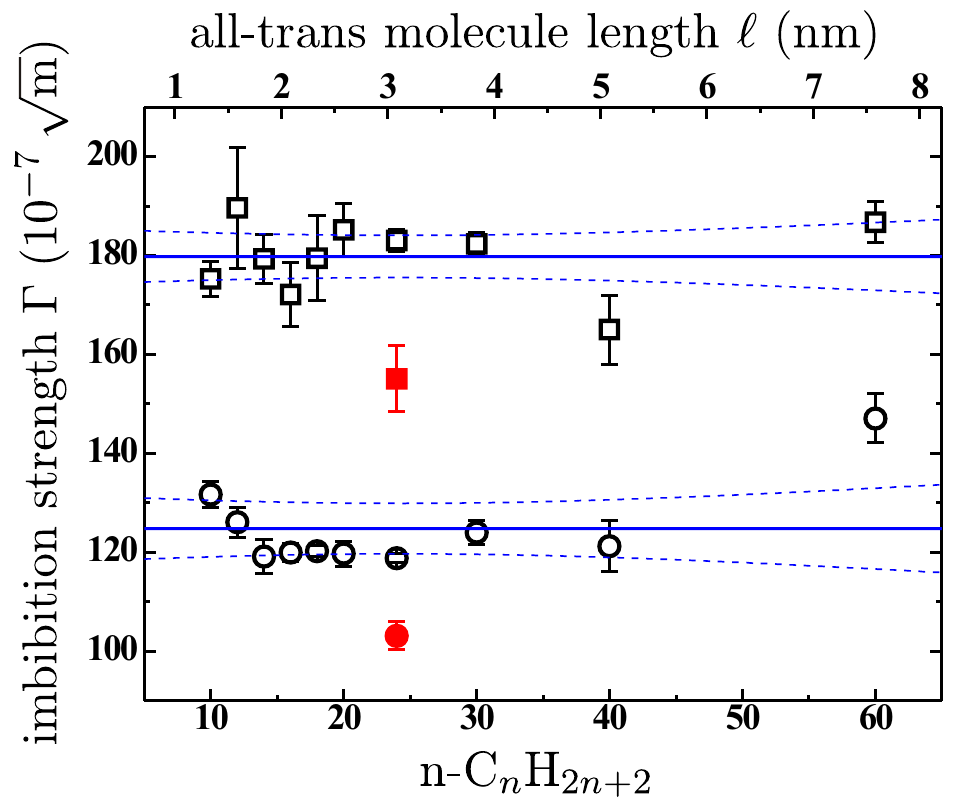}
\caption[Imbibition strengths of hydrocarbons in porous \V deduced from gravimetric measurements]{Imbibition strengths $\Gamma$ obtained from gravimetric measurements for a series of linear hydrocarbons (open symbols) ranging from n-decane up to n-hexacontane invading V10 ($\square$) and V5 ($\fullmoon$), respectively. Their mean values are visualized by the solid lines, lying within the confidence interval (dashed lines, confidence level: 90\,\%). The corresponding all-trans molecule length $\ell$ of the alkane can be read from the top axis. The measured branched hydrocarbon squalane is indicated by the filled symbols.}
\label{alk_conductance}
\end{figure}
Eleven linear hydrocarbons ranging from n-decane (n-C$_{10}$H$_{22}$) to n-hexacontane (n-C$_{60}$H$_{122}$) being tantamount to all-trans molecule lengths $\ell$ between 1\,nm and approximately 8\,nm were applied. Furthermore, the branched hydrocarbon squalane (2,6,10,15,19,23-hexamethyltetracosane) with six additional methyl side groups as compared to n-tetracosane (see \bild{alkanemolecules}) was used. For better statistics the imbibition dynamics of each liquid/substrate combination were ascertained at least three times. The results of the study are summarized in \bild{alk_conductance}. What is more, the invasion of n-hexadecane (n-C$_{16}$H$_{34}$) in V5 was also recorded as a function of the relative humidity (RH) of the surrounding atmosphere. For this purpose the whole setup was put into a box. Applying a humidifier the water content of the atmosphere within this housing could be increased to a maximum RH of approximately 50\,\%. The resultant imbibition strengths $\Gamma$ are listed in \tab{tab:GammaGravHum_Alk} together with the extracted values of the initial porosities.

Obviously the invasion dynamics are suitably described through the arithmetic averages over all measured n-alkanes $\Gamma = (122.8\pm 2.9)\cdot 10^{-7}\,\sqrt{\rm m}$ for V5 and $\Gamma = (178.8\pm 2.5) \cdot 10^{-7}\,\sqrt{\rm m}$ for V10. No distinct influence of the all-trans chain length $\ell$ on the respective n-alkane's invasion dynamics can be detected. Only for the largest molecule n-hexacontane ($\ell\approx 7.6$\,nm) in the smaller pores of V5 ($\subs{r}{0}= 3.4 $\,nm) hints for significant deviations from the arithmetic average towards increased overall imbibition dynamics are found $\left[\Gamma = (147.1\pm 5.0) \cdot 10^{-7}\,\sqrt{\rm m}\right]$. Interestingly, enhanced dynamics for long-chain molecules (mostly polymers) have already been documented in the literature \cite{Binder07, Priezjev04, Shin07} and could be ascribed to substantial slip at the liquid-substrate interface. This possibility will be addressed later on.

Again, the hydrodynamic pore radii $\subs{r}{h}$ can be deduced from the measured imbibition strengths $\Gamma$. One attains $\subs{r}{h} = (2.85\pm 0.22)$\,nm for V5 and $\subs{r}{h} = ( 4.49 \pm  0.28 )$\,nm for V10, or in terms of the slip length $b = \subs{r}{h}-\subs{r}{0}$
\begin{eqnarray}
b & \approx & -5 \,\text{\AA} \qquad  \mathrm{for\;\, linear\;\, hydrocarbons\;\, invading\;\, V5\;\, and}  \nonumber \\   
b & \approx & -4 \,\text{\AA} \qquad  \mathrm{for\;\, linear\;\, hydrocarbons\;\, invading\;\, V10.} \nonumber
\label{eq:sliplengthAlkanes}
\end{eqnarray}
Comparable to the water measurements for both V5 and V10 a {\it negative} slip length of approximately 5\,\AA\ is found.

In this context it is worthwhile considering the impact of the pore walls' initial water coating on the liquid invasion. According to the results presented in \tab{tab:GammaGravHum_Alk} there is no significant influence of the RH on the overall rise dynamics, provided one takes into account the initial porosity as a function of the RH. Obviously the degree of the initial water coverage does not influence the hydrocarbon invasion. This result is especially important regarding Ref.~\cite{Dorris81}, in which already the presence of one to two water monolayers preadsorbed on a silica surface was found to be sufficient to considerably reduce the London force field of the glass. Nevertheless, the measured invasion dynamics prove that the modified alkane-substrate interactions do not change the wetting behavior of the hydrocarbon. 

However, especially the first adsorbed water layer is highly stabilized by the attractive potential between silica and water \cite{Wallacher97}. In addition to that, a water coating on the pore walls is favored compared to an alkane coating. This is expressed by their different surface tensions and can be demonstrated by displacement measurements. Here, an initially full hydrocarbon filling in porous \V is replaced by a water filling through imbibition (after connecting the sample to a water reservoir). Cross checks of the sample's mass change with the different liquid densities reveal an absolutely complete substitution of water for the alkane.

One has to assume that the imbibed hydrocarbon is not able to entirely displace the initial water coverage. At least the first layer adjacent to the pore walls should be present during the capillary rise process. This extremely immobile layer provides a first contribution of $\sim 0.25$\,nm to the overall sticking layer thickness stated above. But there is still a residual of $b$, which can be ascribed to either a second water layer or to a layer of flat lying hydrocarbon molecules. Thanks to their similar thicknesses one cannot distinguish between these two cases beforehand.
  
\begin{table}[!t]
\centering
\setlength\extrarowheight{2pt}
\caption[Imbibition strengths $\Gamma$ and initial porosities $\subs{\phi}{i}$ of n-hexadecane invading V5 as a function of the relative humidity]{Imbibition strengths $\Gamma$ and initial porosities $\subs{\phi}{i}$ of n-hexadecane invading V5 as a function of the relative humidity (RH).} 
\begin{tabularx}{1\textwidth}{>{\centering\arraybackslash}X>{\centering\arraybackslash}X>{\centering\arraybackslash}X} \toprule
RH (\%) & $\phi_{\rm i}$ & $\Gamma$ ($10^{-7}\,\sqrt{\rm m}$)   \\ \midrule \midrule 
                   24   &  0.275    & $116.5\pm 3.8$          \\  
                   34   &  0.262    & $121.9\pm 5.3$        \\                     
                   50   &  0.245    & $121.9\pm 4.1$            \\  \bottomrule                  
\end{tabularx}
\label{tab:GammaGravHum_Alk}
\end{table}

Yet, the concept of a layer of flat lying hydrocarbons is confirmed by the squalane measurements, which show decreased overall invasion dynamics. The $\sim 15$\,\% decrease in the imbibition strength and the consequent additional reduction of the hydrodynamic pore radii [$\subs{r}{h} = (2.61\pm 0.21)$\,nm for V5 and $ \subs{r}{h} = ( 4.19 \pm  0.32 )$\,nm for V10] by $\sim 0.3 $\,nm can uniformly be explained by the larger diameter of the squalane molecule as compared with a non-branched hydrocarbon chain. The extra methyl groups attached to the n-tetracosane backbone cause an increase of the sticking layer's thickness by a value comparable to the decrease in $\subs{r}{h}$ just mentioned. Interestingly, a similar influence of the additional side branches on the boundary conditions was already reported \cite{Schmatko05}. The continuously positive slip lengths $b$ for both n-hexadecane and squalane revealed within this FRAP study notwithstanding, $b$ was found to be always lower for the branched squalane. 

Summarizing, one may assume the sticking layer to be constituted of a water layer directly adjacent to the pore walls and a contiguous layer of flat lying hydrocarbon chains whereas the innermost liquid compartment obeys classical hydrodynamics based on continuum mechanical theory. Not until the chain length of the hydrocarbon reaches the diameter of the pore the results are seemingly influenced by the respective shape of the liquids' building blocks.

The assumption of a sticky monolayer of hydrocarbons is confirmed by forced imbibition experiments on n-alkanes in \V \cite{Debye59}, by studies regarding the thinning of n-alkane films in the surface force apparatus \cite{Chan85, Christenson82, Israelachvili86, Georges93} and by molecular dynamics simulations \cite{Heinbuch89, Cui01}. On top of this, X-ray reflectivity studies indicate strongly adsorbed, flat lying monolayers of hydrocarbons on silica \cite{Mo05, Basu07, delCampo09, Lu2010}. The retained viscosity of the confined alkanes was also verified by means of MD simulations \cite{Gupta97}. 

With that the just stated results disagree with a series of recent publications. Especially the existence of a {\it positive} slip length $b$ of short-chain alkanes was proven several times by experiment \cite{Pit00, Schmatko05} as well as by simulation \cite{Stevens97, Gupta97}, even for complete wetting and hence highly attractive liquid-substrate interaction. Interestingly, the hydrodynamic pore radius $\subs{r}{h} = (3.12 \pm 0.26)\,$nm of n-hexacontane in V5 reveals a slip length $b=-0.28$\,nm, which is, admittedly, still negative. Nonetheless, it is markedly increased as compared with $b=-0.55$\,nm for the arithmetic average over all n-alkanes by approximately 3\,\AA\ -- the thickness of the layer of flat lying hydrocarbons. This result might be interpreted as a step towards a slip boundary condition. Yet, because of the relatively large error bars it must be treated as a mere indication of altered boundary conditions.

\section[Spontaneous imbibition of silicon oils in porous Vycor]{Spontaneous imbibition of silicon oils in porous \Vend}
\begin{figure}[!t]
\centering
\includegraphics*[width=.4 \linewidth]{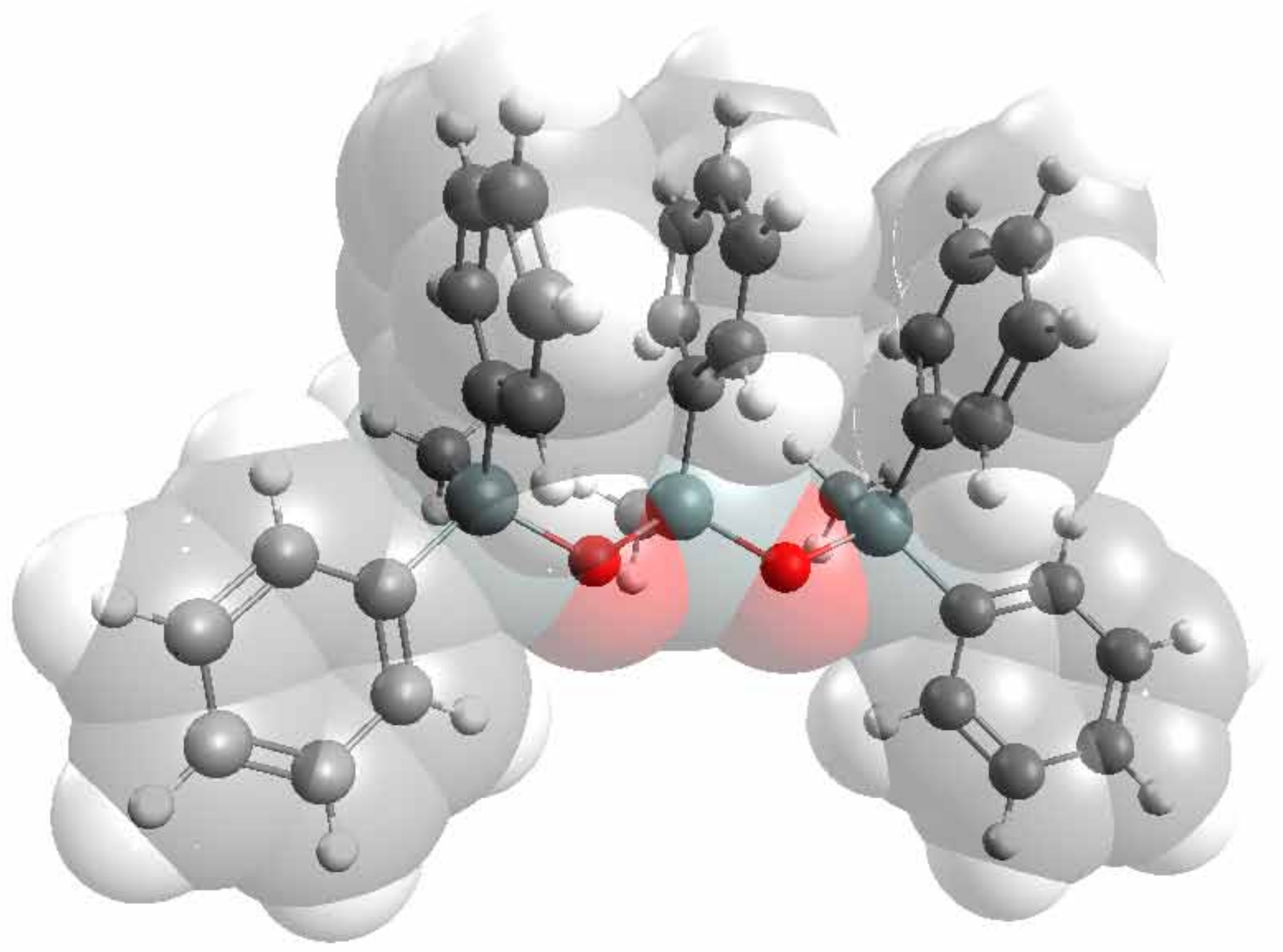}
\caption[Illustration of the molecular structure of the main component of Dow Corning diffusion pump oils]{Illustration of the molecular structure of 1,1,3,5,5-Pentaphenyl-1,3,5-trimethyltrisiloxane. It is the main constituent of the applied Dow Corning diffusion pump oils DC704 and DC705. Five phenyl and three methyl groups are regularly attached to the trisiloxane (\ce{Si3O2-R8}) backbone. The molecular dimensions can roughly be estimated using the diameter of a phenyl ring ($\sim 5.4$\,\AA) to be between 1\,nm and 1.5\,nm.}
\label{DC705}
\end{figure}
So far the results on the flow of water and linear hydrocarbons in extreme spatial confinement revealed a retained fluidity and capillarity of the liquids. Considering the highly attractive interaction between fluid and substrate the verification of a sticking layer is not too surprising. Liquids slipping at a substrate are mainly observed in non-wetting configurations \cite{Vinogradova99, Lasne08, Fetzer07, Barrat99}. Unfortunately such systems cannot be investigated by means of spontaneous imbibition, which always requires a wetting liquid. 

However, applying polymeric or oily liquids also seems to facilitate the occurrence of slip at the fluid-solid interface \cite{Priezjev04}. This presumption is corroborated by a recent imbibition study of polystyrene \cite{Shin07}. What is more, the MD simulations carried out in Ref.~\cite{Binder07} demonstrate a slip boundary condition for the capillary rise of a short-chain macromolecule (representative of the building blocks of silicon oil) with slip lengths on the order of 30\,\% of the tube radius. Therefore, when one tries to observe modified boundary conditions it will be rather promising to use such liquids. 
\begin{table}[!b]
\centering
\setlength\extrarowheight{2pt}
\caption[Imbibition strengths and slip lengths of si\-li\-con oils in po\-rous \V deduced from gravimetric measurements]{Characterization of the imbibition dynamics of Dow Corning silicon oils in porous \V at room temperature by means of the imbibition strength $\Gamma$ (in $10^{-7}\,\sqrt{\rm m}$) and the resultant slip length $b$ (in nm).}
\begin{tabularx}{1\textwidth}{>{\centering\arraybackslash}X>{\centering\arraybackslash}X>{\centering\arraybackslash}X>{\centering\arraybackslash}X>{\centering\arraybackslash}X} \toprule
\multirow{2}{*}{DC\ldots} & \multicolumn{2}{c}{V5} & \multicolumn{2}{c}{V10}           \\ \cmidrule{2-5}
       & $\Gamma$  & $b$                & $\Gamma$ & $b$ 						\\ \midrule \midrule
  704  & $63.6\pm 5.6$       &  $-1.35\pm 0.23$   & $109.2\pm 12.6$    & $-1.39\pm 0.40$  \\
  705  & $53.2\pm 4.8$       &  $-1.52\pm 0.21$   & $122.4\pm 11.4$    & $-1.19\pm 0.38$  \\ \bottomrule
\end{tabularx}
\label{tab:GammaM_DCoils}
\end{table}

For this purpose we applied the diffusion pump oils DC704 and DC705 from Dow Corning. Their main constituent is the siloxane depicted in \bild{DC705} with molecular dimensions between 1\,nm to 1.5\,nm. They are characterized by a negligible vapor pressure and relatively high viscosities up to 200\,mPa$\cdot$s. Because of the latter the capillary rise process is substantially slowed down. The BCLW-$\sqrt{t}$ behavior can be observed for weeks. Again, this preeminently elucidates the retained fluidity and Newtonian character of the liquid. The extracted imbibition strengths $\Gamma$ are listed in \tab{tab:GammaM_DCoils}. 

Compared to the values of water or the hydrocarbons the dynamics of the silicon oils are decreased. This is equivalently expressed in terms of the resultant slip lengths $b$ (see \tab{tab:GammaM_DCoils}). They all indicate a sticking layer boundary condition whose thickness is approximately 1.4\,nm. Interestingly this result is independent of both the sample type and the silicon oil. Thus it is not too erroneous to conclude that, again, one layer of siloxane molecules is pinned to the pore walls and does thereby substantially lower the invasion dynamics. 

This result does not necessarily have to contradict the findings of the studies mentioned at the beginning \cite{Binder07, Priezjev04, Shin07}. The applied silicon oils presumably are not chain-like enough but rather block-like. Possibly this internal molecular structure is decisive for the occurrence of slip at the interface. In this regard one has to recall the increased dynamics of n-hexacontane in V5 compared to the short-chain hydrocarbons. This result stimulates future studies on more extended alkane and polymer melts with molecular weights beyond 1\,$\frac{\rm kg}{\rm mol}$.

\section{Phase transitions of liquids in nanopores explored by spontaneous imbibition}

In this final section the influence of the nanopore confinement on temperature induced phase transitions will be examined. Usually phase transitions are accompanied by unique variations of the fluid properties. Since the overall rise dynamics sensitively depend on these quantities one can easily detect such characteristic deviations. According to \rel{eq:imb_mt} the desired information is comprised in the prefactor $\subs{C}{m}$ of the observed $\sqrt{t}$ behavior of the measured mass increase:
\begin{equation}
{ \subs{C}{m}} = {\rho}\, \sqrt{\frac{\sigma}{\eta}}\; \cdot\; A \,\sqrt{\frac{\subs{\phi}{i}\,\cos{ \subs{\theta}{0}}}{2}}\,\Gamma \; ,
\label{eq:Cm}
\end{equation}
with only the quantities in the first factor depending on the temperature. Here, the slight $T$ dependence of $\subs{\phi}{i}$ can be neglected in the limited $T$ regions observed. Moreover, the static contact angle $\subs{\theta}{0}$ is zero for all $T$. The matrix properties are $T$-invariant and can be eliminated through normalization. Hence, by performing several imbibition measurements at different temperatures in the $T$-range of interest a simple comparison of the prefactors might already provide an insight into the phase transition behavior of the confined liquid.

\subsection[Meniscus freezing in n-tetracosane]{Meniscus freezing in n-tetracosane} \label{sectionSF}
At first we present a $T$-dependent study of the capillary rise of the linear hydrocarbon n-tetracosane (n-C$_{24}$H$_{50}$) in porous \Vend. The chosen alkane exhibits a surface freezing transition\label{surfacefreezingtemperature} at $\subs{T}{s}=54$\dC. 
% which is accompanied by a characteristic reversal of the $\sigma(T)$-slope. This distinctive behavior must affect the rise dynamics, which permits one to examine the phase transition in nanopore confinement.
Surface freezing is the formation of a single solid monolayer floating on the bulk melt in a $T$-range between the bulk freezing temperature $\subs{T}{f}$\label{bulkfreezing} and a temperature $\subs{T}{s}$ \cite{Earnshaw92, Ocko97, Sirota97, Wu93a, Wu93}. Upon surface freezing the hydrocarbon molecules are rectified, parallel-aligned with their long axis along the surface normal, and the center of mass lattice is hexagonal, resulting in a 3\,nm thick surfactant-like monolayer. The abrupt onset of surface freezing at $\subs{T}{s}$ allows one to switch on (and off) the 2D crystalline phase at the free surface by a small $T$ variation. The goal of this study is to investigate whether this particular molecular rearrangement is detectable by and how it affects the imbibition dynamics of the alkane in nanopores. %This phase transition is known from a series of n-alkanes ranging from n-C$_{14}$H$_{30}$ to n-C$_{50}$H$_{102}$, however, we selected n-tetracosane for this study, since it exhibits the largest surface freezing $T$-range ($\sim 3$\dC) of all pure n-alkanes.

%\begin{figure}[!t]
%\centering
%\includegraphics*[width=.7 \linewidth]{surface_freezing}
%\caption[Illustration of the surface freezing transition]{Illustration of the surface freezing transition. Between the bulk melting point $\subs{T}{f}$ and the phase transition temperature $\subs{T}{s}$ one single solid monolayer is established on the free surface of the liquid.}
%\label{surface_freezing}
%\end{figure} 
\begin{figure}[!b]
\centering
\includegraphics*[width=.6 \linewidth]{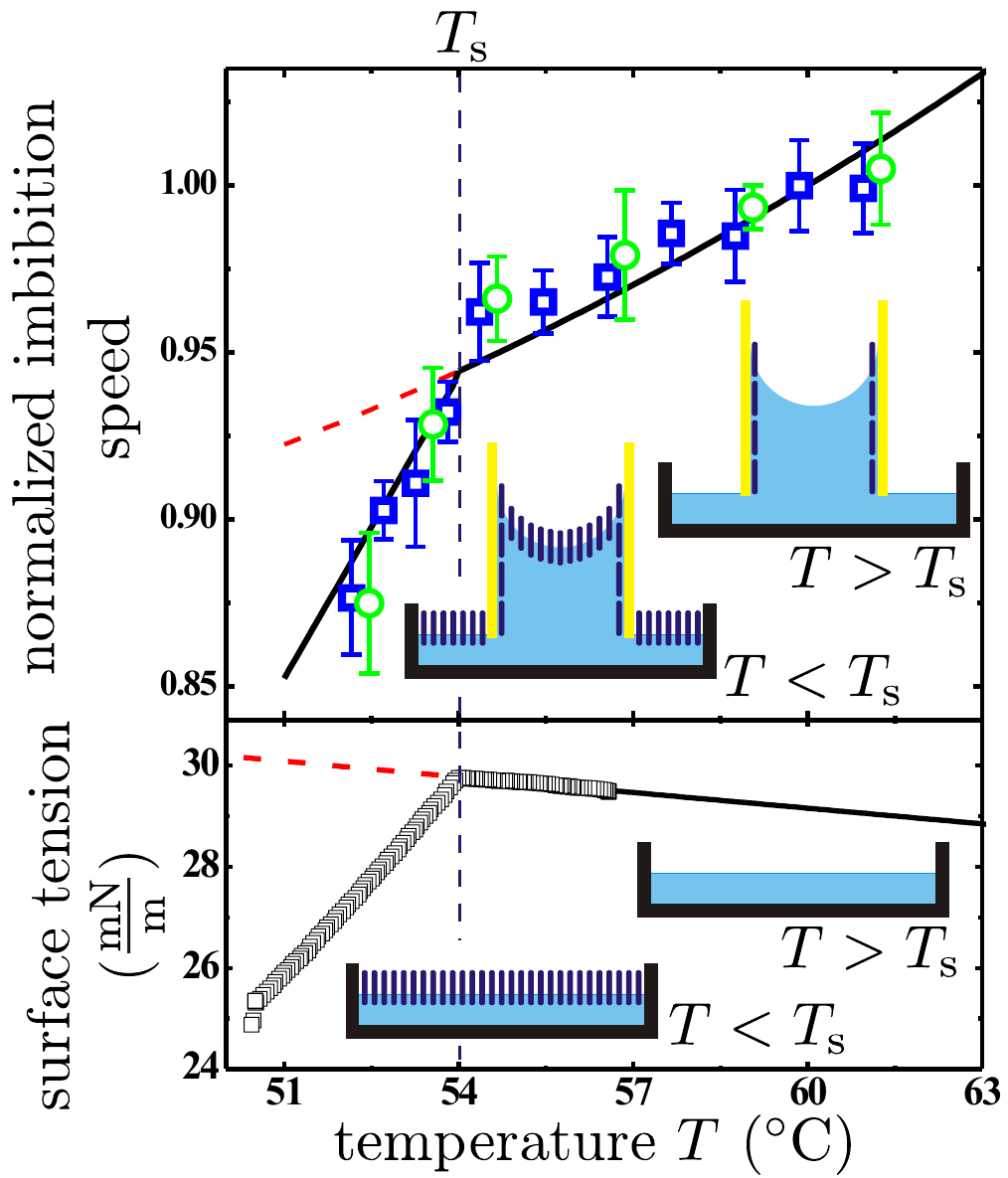}
\caption[Measured normalized imbibition speeds of tetracosane in V5 and V10 as a function of the temperature]{(upper panel): Measured normalized imbibition speeds $v^{\rm n}$ (for $T^{\rm n}=60$\dC) of tetracosane in V5 ({\color{blue} $\square$}) and V10 (${\color{green} \fullmoon}$), respectively, in comparison to values calculated under the assumption of meniscus freezing (\Flatsteel). Insets: Illustration of a surface frozen layer at an advancing meniscus ($T<T_{\rm s}$) and of a sticky, flat lying boundary layer at the silica pore wall in the entire $T$-range investigated. (lower panel): $T$-dependent surface tension of tetracosane. Insets: Illustrations of surface freezing at a planar tetracosane surface. The dashed lines correspond to extrapolations of the calculated imbibition speeds and of the surface tension below $T_{\rm s}$ in the absence of surface freezing, respectively.}
\label{SF_speeds}
\end{figure} 

% The surface tension $\sigma$ is a direct measure of the surface excess free energy: 
%\begin{equation}
%\sigma (T) = \subs{\epsilon}{s} - \subs{\epsilon}{b} - T\cdot (\subs{S}{s} - \subs{S}{b})
%\label{eq:sigmaT}
%\end{equation}
%where $\subs{\epsilon}{s}$ and $\subs{\epsilon}{b}$ are the energies and $\subs{S}{s}$ and $\subs{S}{b}$ the entropies for the surface and the bulk phase, respectively. The temperature slope of surface tension shows information on the surface excess entropy: ${\rm d}\sigma/{\rm d}T=-(S_{\rm s}-S_{\rm b})$. The negative slope of bulk tetracosane for $T>T_{\rm s}$ is typical of an ordinary liquid surface, for which the molecules on the surfaces are less constrained than those in the bulk, thus $S_{\rm s}$ is slightly larger than $S_{\rm b}$, yielding ${\rm d}\sigma/{\rm d}T<0$. Surface freezing and its first-order character result in an abrupt reduction of the surface entropy $S_{\rm s}$ in such a way that $S_{\rm s}$ is smaller than $S_{\rm b}$ leading to ${\rm d}\sigma/{\rm d}T>0$. 

The surface freezing transition is accompanied by a distinctive signature in the $T$ behavior of the surface tension $\sigma$, known from bulk n-alkanes, that is a change from a small negative, above $\subs{T}{s}$, to a large positive $T$-slope, below $\subs{T}{s}$ (see lower panel in \bild{SF_speeds}) \cite{Ocko97}. This distinctive behavior of the surface tension must be mapped in the $T$-de\-pen\-dence of the measured mass uptake. 

For convenience the measured prefactors $\subs{C}{m} (T)$ of the $\sqrt{t}$-imbibition dynamics are normalized by the prefactor value at an arbitrarily chosen temperature $T^{\rm n}$. This procedure finally yields the normalized imbibition speed\label{normimbspeed} $v^{\rm n}$ of the measurement at the temperature $T$. One should not be confused by the term `speed', since $v^{\rm n}$ is actually dimensionless. Nevertheless, it quantifies the mass uptake rate of the sample at a given temperature. According to \rel{eq:Cm} this quantity can also be calculated:
\begin{equation}
v^{\rm n}(T) \equiv \frac{\subs{C}{m} (T)}{\subs{C}{m} (T^{\rm n})} = \frac{\rho(T)}{\rho(T^{\rm n})} \cdot \sqrt{\frac{\sigma(T) \cdot \eta(T^{\rm n})}{\sigma(T^{\rm n}) \cdot \eta(T)}} \; .
\label{eq:normimbspeed}
\end{equation}
It is important to note that in \rel{eq:normimbspeed} the geometry as well as the internal topology of the substrate do not play a role any longer. Assuming the $\sigma(T)$ kink-behavior according to \bild{SF_speeds} along with the $T$ dependency of $\eta$ and $\rho$ one can calculate theoretical values of $v^{\rm n}(T)$. They are plotted as a solid line in the upper panel of \bild{SF_speeds}. The measured speeds are indicated by the single points. Due to the remarkable coincidence of the temperature characteristic of the kink in $v^{\rm n}$ and the degree of deviation from a suppressed phase transition we feel encouraged to solely attribute the distinct change in the imbibition dynamics to a change in $\sigma(T)$ at the advancing menisci typical of surface freezing. The corresponding nanoscopic flow configuration is illustrated in the insets of \bild{SF_speeds} for both $T<\subs{T}{s}$ and $T>\subs{T}{s}$. We would refer the interested reader to a more detailed publication reporting on the surface freezing of n-tetracosane in nanoporous silica \cite{Gruener09a}.

\subsection[Flow of a rod-like liquid crystal (8OCB) in Vycor]{Flow of a rod-like liquid crystal (8OCB) in Vycor}
Finally, the phase transition behavior of a thermotropic liquid crystal in nanopore confinement was examined. For this purpose we applied one of the best scientifically studied liquid crystals, the rod-like octyloxycyanobiphenyl (8OCB, see \bild{8OCB} for an illustration of its molecular structure). At $\sim 80$\dC\ it undergoes a transition from the isotropic to a nematic phase, which is accompanied by the occurrence of a characteristic shear viscosity minimum. Again, this distinctive behavior must affect the rise dynamics which enables one to examine the phase transition in nanopore confinement. The most common phases are illustrated in \bild{LC_phases}.

%Basically, liquid crystals (LCs) are anisotropic liquids. They possess the fluidity of a true liquid, as well as varying degrees of long range order normally associated with crystalline solids. this intrinsic constitution of the liquid gives rise to the alignment of the molecules and, therefore, to the anisotropic properties of the liquid crystal. 
\begin{figure}[!t]
\centering
\includegraphics*[width=.5\linewidth]{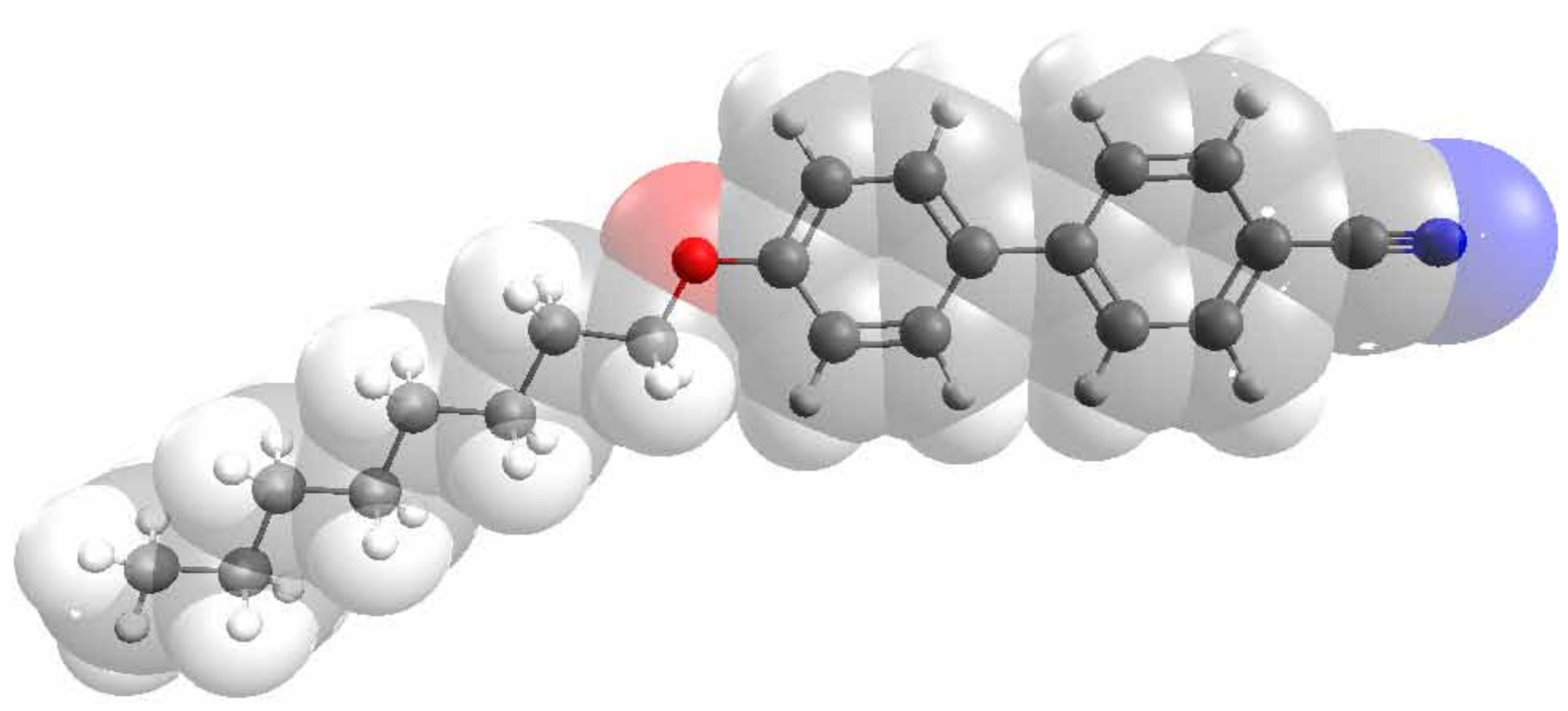}
\caption[Illustration of the molecular structure of the liquid crystal 8OCB]{Illustration of the rod-like molecular structure of the liquid crystal octyloxycyanobiphenyl (8OCB) with the phase transition behavior: crystalline $\stackrel{54\,^{\circ}{\rm C}}{\longrightarrow}$ A $\stackrel{67\,^{\circ}{\rm C}}{\longrightarrow}$ N $\stackrel{80\,^{\circ}{\rm C}}{\longrightarrow}$ I. The molecule's head (comprising the cyano and the biphenyl group) and its tail (the octyl chain) each have a length of slightly more than 1\,nm. Its overall length is $\sim 2$\,nm. As opposed to the hydrocarbons the 8OCB molecule is rather rigid.}
\label{8OCB}
\end{figure}

%Liquid crystal phases are called mesophases. Thermotropic liquid crystals exhibit one or more anisotropic liquid phases between the melting point $\subs{T}{f}$ and the temperature at that they become isotropic $\subs{T}{c}$; this temperature is called the clearing point\label{clearingpoint} due to the milky, translucent appearance of liquid crystal phases. These phases generally occur reproducible with heating or cooling.
\begin{figure}[!t]
\centering
\includegraphics*[width=.7 \linewidth]{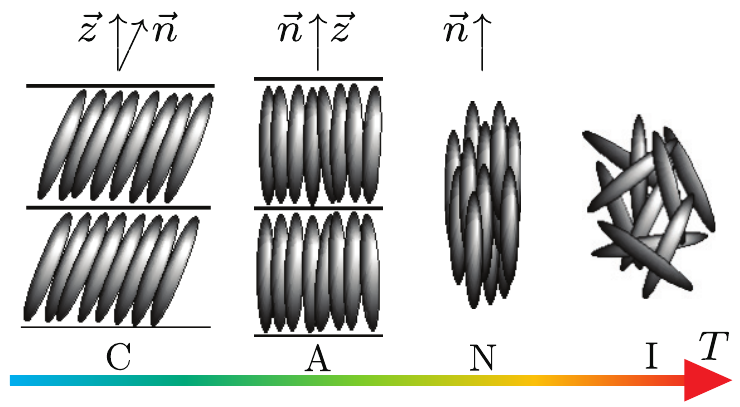}
\caption[Illustration of common liquid crystal phases]{Illustration of common liquid crystal phases. Upon heating a frozen liquid crystal one might pass the following mesophases: smectic C (C), smectic A (A), and nematic (N). Beyond the clearing point $\subs{T}{c}$ the liquid crystal is in the isotropic state (I).}
\label{LC_phases}
\end{figure}

%In the isotropic (I) state the rod-shaped molecules are randomly oriented and the molecular centers of mass move as in any liquid. Upon cooling the molecules, at a certain well defined temperature, often self-reorient with their long axes parallel to each other whereas the centers of mass are still isotropically distributed. This phase is called the nematic phase (N). For an isotropic liquid, averaging molecular orientations gives no result, since there are as many molecules lying along one axis as along another. In the nematic phase, averaging molecular orientations gives a definite preferred direction, which is referred to as the director $\vec{n}$. However, it is important to remember that liquid crystals are liquids, meaning that, although there is an average order, molecules are constantly flowing and moving, changing position and orientation.

%Upon further cooling from the nematic phase, the molecules will often self-assemble into layers. These layered phases are called smectic liquid crystals. When the director is parallel to the layer normal $\vec{z}$, the phase is referred to as a smectic A (A). When cooling from the A phase, a more ordered smectic phase is often seen, in which the molecules are tilted with respect to the layers. This tilted, layered phase is called the smectic C (C). 

\subsubsection{Nemato-Hydrodynamics: Shear viscosity minimum and presmectic divergence of flowing nematic liquid crystals}
The momentum transport in nematic liquid crystals shows an anisotropy since it depends on the mutual orientations of the macroscopic molecular alignment (the director $\vec{n}$), the flow velocity ($\vec{v}$) and the velocity gradient ($\nabla\, v$). In 1946 Mieso\-wicz\label{Miesowicz} defined three principal shear viscosity coefficients of nematics \cite{Miesowicz46}, which can be measured in three different Couette flow experiments sketched in \bild{LC_visc}. Typically magnetic fields are applied in order to align the molecules in the nematic sample. Intuition suggests that the lowest resistance to the nematic flow, i.e. the lowest viscosity value, should be $\eta_2$. Among the two remaining viscosities, $\eta_1$ should have the highest value.
\begin{figure}[!t]
\centering
\includegraphics*[width=.55 \linewidth]{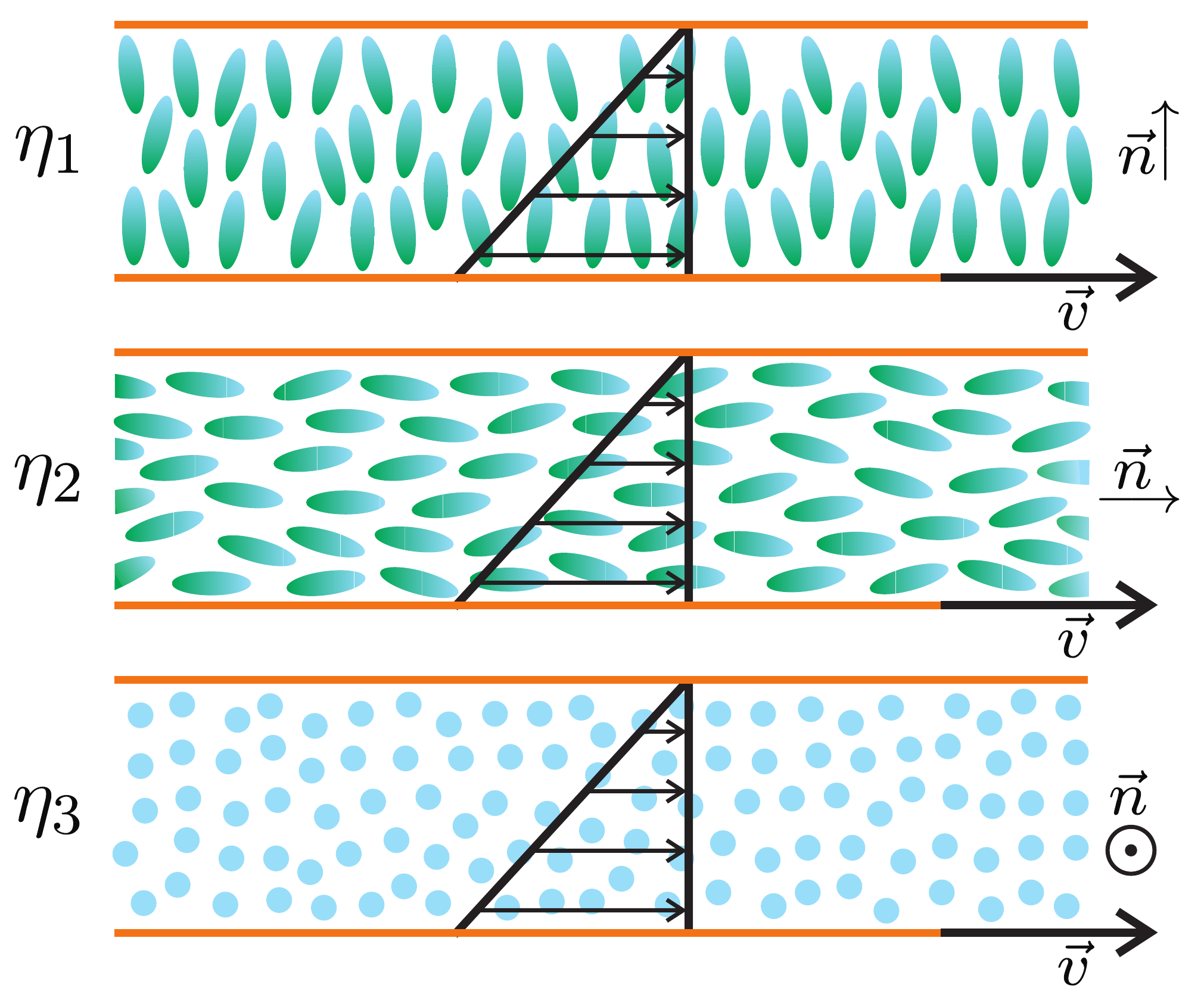}
\caption[Illustration of the three Miesowicz shear viscosity coefficients of nematic liquid crystals]{The experimental Couette flow conditions for measurements of the three Miesowicz shear viscosity coefficients of nematic liquid crystals: $\eta_1$ for $\vec{n}\,\bot\, \vec{v}$ and $\vec{n} \,||\, \nabla \,v$, $\eta_2$ for $\vec{n}\,||\, \vec{v}$ and $\vec{n} \,\bot\, \nabla \,v$, $\eta_3$ for $\vec{n}\,\bot\, \vec{v}$ and $\vec{n}\, \bot\, \nabla \,v$.}
\label{LC_visc}
\end{figure}
\begin{figure}[!t]
\centering
\includegraphics*[width=.55 \linewidth]{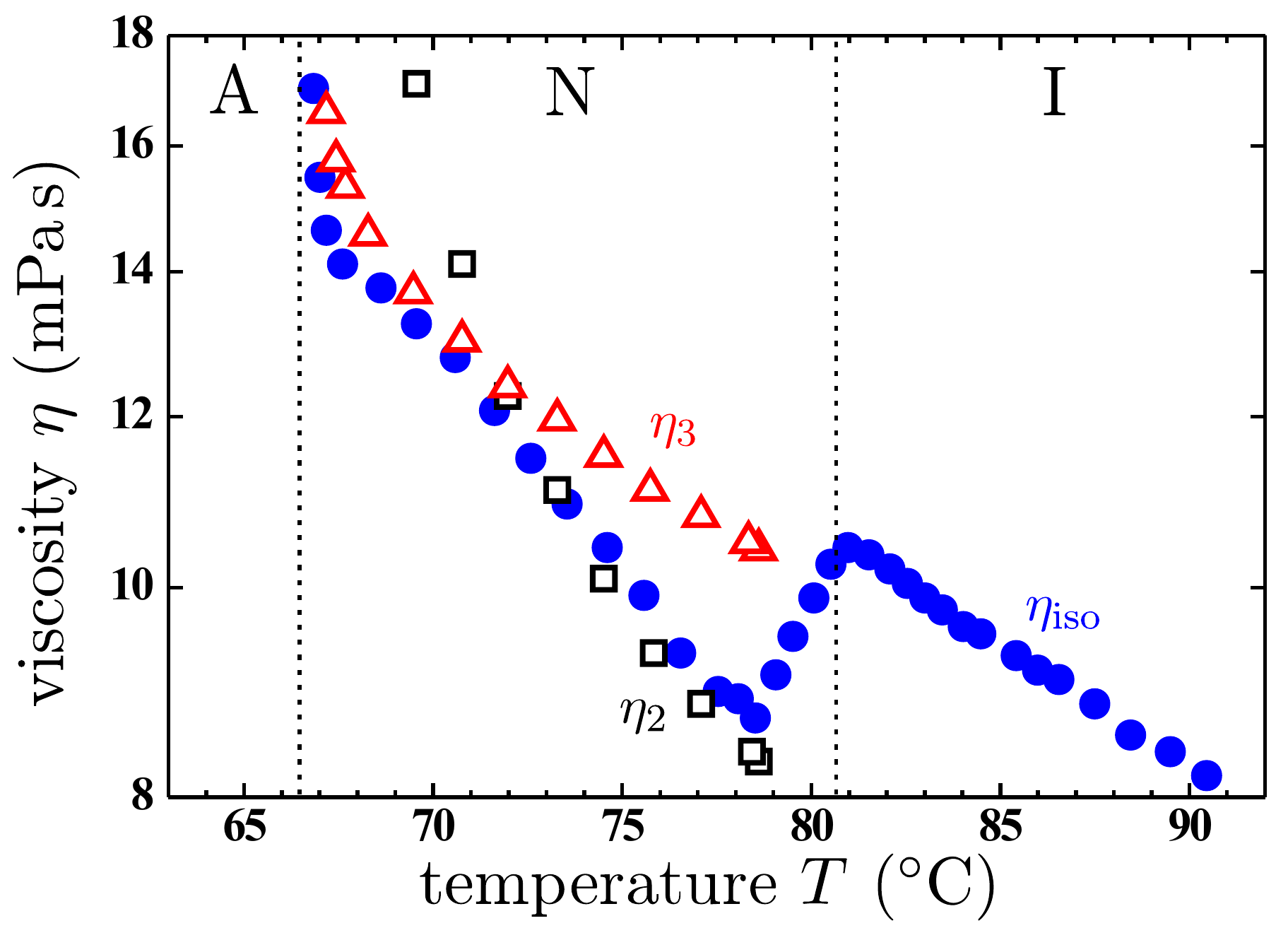}
\caption[Miesowicz and free flow viscosities of 8OCB]{Miesowicz shear viscosities $\eta_2$ ($\square$) and $\eta_3$ (${\color{red} \triangle}$) of the liquid crystal 8OCB compared to its free flow viscosity $\subs{\eta}{iso}$ ({\color{blue} \Circsteel}) according to \cite{Graf92}.}
\label{LC_visc_8OCB}
\end{figure}

Indeed, when the orienting magnetic field, i.e. the director $\vec{n}$, is parallel to the velocity $\vec{v}$ of the nematic flow, the lowest viscosity is recorded (see open symbols in \bild{LC_visc_8OCB}). Nevertheless, this relatively simple picture of the viscosity of nematic liquid crystals is disturbed for the compounds exhibiting the transition to the smectic A phase. Then, with decreasing temperature, the viscosity $\eta_2$ shows a strong increase and goes to infinity at the N-A phase transition. The $\eta_1$ and $\eta_3$ viscosities are almost unaffected. At a temperature that is a few degrees below the temperature at which the transition to the smectic A phase takes place, the viscosities $\eta_2$ and $\eta_3$ interchange their roles and then the lowest nematic viscosity corresponds to the flow in the $\eta_3$ configuration. The presmectic behavior of the $\eta_2$ viscosity is due to the formation of precursors of smectic planes with $\vec{z}\,||\,\vec{v}$ that would be immediately destroyed by the velocity gradient, thus this configuration is rendered unfavorable.

The behavior of the freely flowing compound obeys a general principle that can be formulated as follows: a free fluid adopts such a manner of flow, as corresponds to the minimum of its viscosity at given conditions \cite{Jadzyn01}. Accordingly, the transition from the isotropic to the nematic phase manifests itself in a strong decrease of the shear viscosity $\subs{\eta}{iso}$ that is very close to $\eta_2$. Consistently, beyond the presmectic cross-over of $\eta_2$ and $\eta_3$ the viscosity of the freely flowing liquid crystal $\subs{\eta}{iso}$ follows $\eta_3$ (see filled symbols in \bild{LC_visc_8OCB}). This result is interpreted in terms of rearrangements of the molecular alignment $\vec{n}$ with respect to the velocity field $\vec{v}$, which can easily be assessed by means of examinations of the compound's viscosity.

This distinctive behavior of the viscosity must be mapped in the $T$-dependence of the measured mass uptake. In \bild{LC_massincrease} some representative measurements of 8OCB invading V5 are shown along with their corresponding $\sqrt{t}$-fits. In analogy to the procedure employed for the evaluation of the surface freezing transition we conducted a quantitative analysis of the phase transition behavior of 8OCB in nanopore confinement referring to the prefactors obtained from the fits.
\begin{figure}[!b]
\centering
\includegraphics*[width=.6 \linewidth]{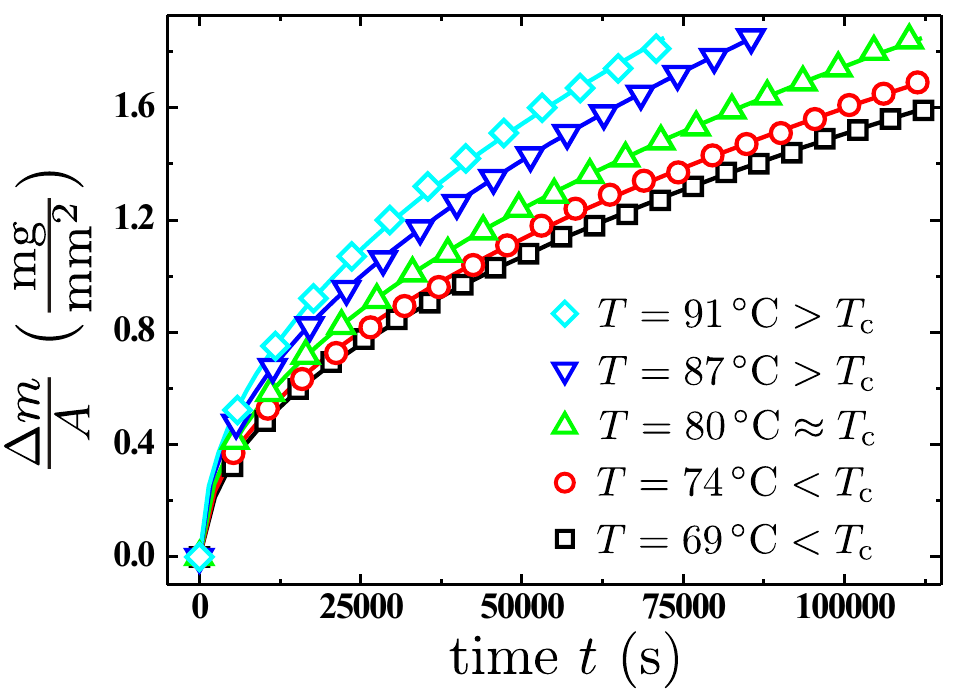}
\caption[Mass uptake of V5 due to liquid crystal imbibition at selected temperatures]{Specific mass uptake of V5 due to the imbibition of the liquid crystal 8OCB as a function of the time for selected temperatures below and above the clearing point $\subs{T}{c} \approx 80$\dC. Solid lines correspond to $\sqrt{t}$-fits. The data density is reduced by a factor of 2500.}
\label{LC_massincrease}
\end{figure}

Again, the prefactors are normalized by the prefactor value at an arbitrarily chosen temperature $T^{\rm n}$. The thereby obtained normalized imbibition speeds $v^{\rm n}(T)$ calculated from the fitting parameters are indicated by the single points in the upper panel of \bild{LC_speeds}. Assuming the $\subs{\eta}{iso}(T)$ behavior according to \bild{LC_visc_8OCB} along with the $T$ dependency of $\sigma$ and $\rho$ one can calculate the theoretical values of $v^{\rm n}(T)$ based on \rel{eq:normimbspeed}. They are plotted as a solid line in the upper panel of \bild{LC_speeds}. 
\begin{figure}[!t]
\centering
\includegraphics*[width=.6 \linewidth]{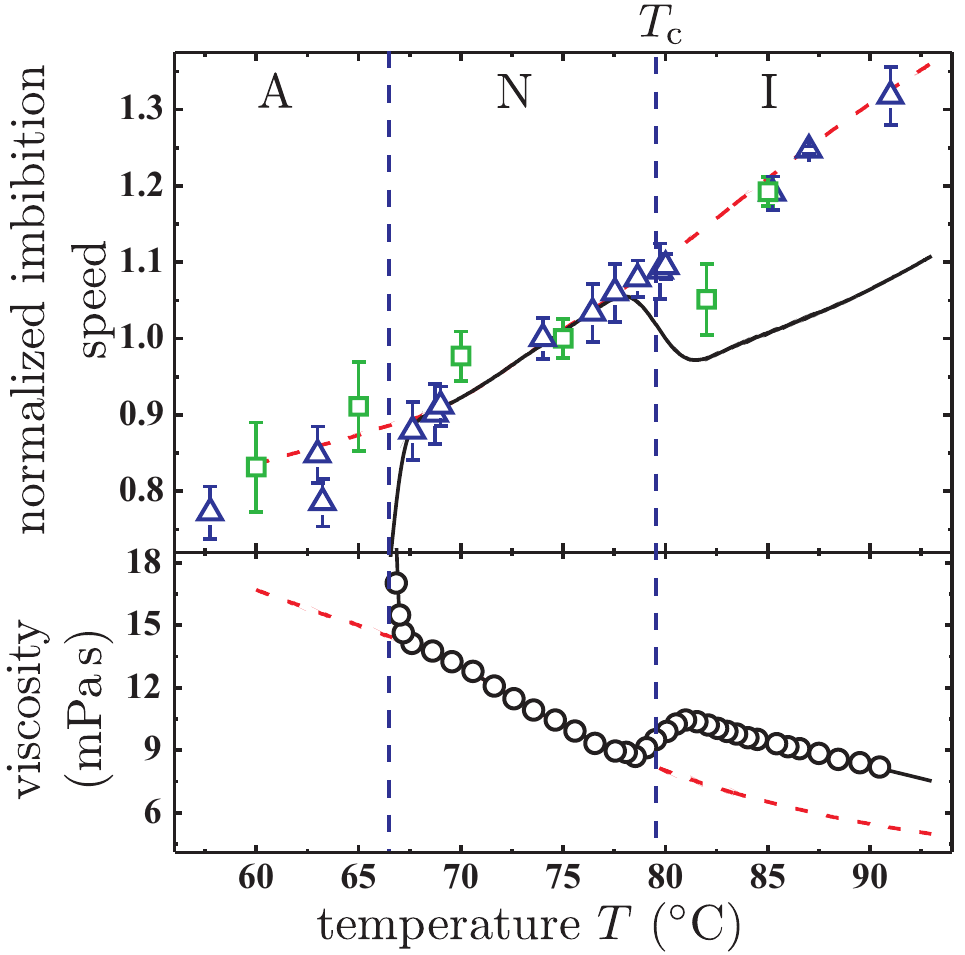}
\caption[Measured normalized imbibition speeds of 8OCB in V5 and V10 as a function of the temperature]{(upper panel): Measured normalized imbibition speeds $v^{\rm n}$ (for $T^{\rm n}=75$\dC) of 8OCB in V5 (${\color{blue} \triangle}$) and V10 ({\color{green} $\square$}), respectively, in comparison with values calculated on the basis of the viscosity values in the lower panel (\Flatsteel). (lower panel): $T$-dependent viscosity of 8OCB by way of comparison (see \bild{LC_visc_8OCB} for a detailed view). The dashed lines correspond to extrapolations of the calculated imbibition speeds and of the viscosity in the absence of the shear viscosity minimum and the presmectic divergence, respectively.}
\label{LC_speeds}
\end{figure} 

The results show a variety of astonishing features. First of all, both V5 and V10 reveal comparable characteristics of the invasion dynamics of the liquid crystal 8OCB. Anyhow, only in the nematic phase the $T$-dependent behavior of the measured imbibition speeds coincides with the prediction based on the $\subs{\eta}{iso}$ values presented in \bild{LC_visc_8OCB}. In particular, the distinctive bump in the proximity of the clearing point $\subs{T}{c}$ as the direct manifestation of the shear viscosity minimum in the theoretical behavior of $v^{\rm n}$ is unambiguously absent. The imbibition speed rather increases monotonously with increasing temperature. 

As mentioned before this distinctive feature of the viscosity at the clearing point is caused by the inset of an alignment of the molecules (in the nematic phase) with respect to the flow direction, that is $\vec{n} \,||\,\vec{v}$. Therefore its absence can intuitively be interpreted in terms of an already existent alignment of the molecules beyond $\subs{T}{c}$. It is obvious to conclude that such an alignment is easily induced by the extreme spatial confinement to cylindrical pores with diameters that are not more than five times the length of the molecule itself. In addition, considering their rigidity and, in particular, the parabolic flow profile established in the pore, it is hard to think of any alternative to the tendency of molecular alignment parallel to the pore axis and, consequently, to the flow velocity $\vec{v}$. From this point of view the absence of the shear viscosity minimum and, consistently, of the nematic to isotropic phase transition is not surprising at all but rather consequent. In this context it is more suitable to label the phase beyond the clearing point not isotropic but {\it paranematic} (P). This term preeminently discloses the solely confinement-induced alignment of the liquid crystal. 

It was demonstrated experimentally \cite{Iannacchione93, Dadmun93, Iannacchione94, Crandall96, Cloutier06}, in agreement with expectations from theory \cite{Sheng76, Steuer04, Cheung06}, that there is no `true' I-N transition for liquid crystals confined in geometries spatially restricted in at least one direction to a few nanometers. The anchoring at the confining walls, quantified by a surface field, imposes a partially orientational, that is, a partially nematic alignment of the confined liquid crystal, even at temperatures $T$ far above the clearing point $\subs{T}{c}$. The symmetry breaking does not occur spontaneously, as characteristic of a genuine phase transition, but is enforced over relevant distances by the interaction with the walls.

The assumption of such a paranemtic phase is corroborated by recent birefringence measurements of liquid crystals confined to an array of parallel, nontortuous channels of 10\,nm mean diameter and 300\,\textmu m length in a monolithic silica membrane \cite{Kytik08}. These measurements elucidate that the surface anchoring fields render the bulk discontinuous I-N transition to a continuous P-N transition. The transition temperature $\subs{T}{c}$ is found to be changed only marginally, due to a balance of its molecular alignment induced upward and its quenched disorder (attributable to wall irregularities) induced downward shift. This agrees with the observations of liquid crystals imbibed in tortuous pore networks \cite{Iannacchione93, Dadmun93}.
%\begin{figure}[!t]
%\centering
%\includegraphics*[width=.70 \linewidth]{LC_birefr_7CB}
%\caption[Optical birefringence of 7CB measured in the bulk state and in silica nanochannels]{Optical birefringence of the liquid crystal 7CB measured in the bulk state (upper panel) and in the silica nanochannels (lower panel) as a function of the temperature $T$ in comparison to fits (solid lines) based on a model discussed in Ref.~\cite{Kytik08}. The finite birefringence characteristic of the paranematic phase is shaded down to the P-N transition temperature. The dashed line marks the bulk clearing point. As insets in the upper and lower panel, the bulk isotropic (I$^{\rm b}$) as well as the bulk nematic (N$^{\rm b}$) phases upon homeotropic alignment, and the confined paranematic (P) and nematic (N) phases are illustrated, respectively. Courtesy of Patrick Huber, Saarland University, Saarbr?cken, Germany.}
%\label{LC_birefr_7CB}
%\end{figure} 

Interestingly, due to the complete absence of the shear viscosity minimum in the results of the $T$-dependent series of imbibition measurements shown in \bild{LC_speeds} a definition of a P-N transition temperature in confinement is not possible at all. This is confirmed by a simple extrapolation of the $\eta_2$ viscosity to higher $T$, which saliently reproduces the measured imbibition speeds beyond the bulk clearing point. This can only be interpreted in terms of an extremely high degree of orientation already existent in the paranematic phase; at least higher than suggested by the birefringence measurements \cite{Kytik08}. This difference is most probably due to the basically differing detection method of the P-N transition: normally birefringence or calorimetry (DSC) measurements performed with the confined {\it static} liquid are applied for this purpose. But, the viscosity measurements presented here refer to the liquid's dynamics in nanopore confinement. As already mentioned before, the additional emerging flow velocity and in particular the velocity gradient seemingly enhance the paranematic orientational alignment in the $\eta_2$-configuration significantly.

The second remarkable feature of the results presented in \bild{LC_speeds} is the absence of the presmectic divergence of the viscosity of the freely flowing liquid crystal. This would result in a dramatic drop of the imbibition speed due to the inset of smectic layering. However, the measured values do not indicate such an effect. Its absence rather suggests a suppression of the A phase in favor of the N phase. This is elucidated by a simple extrapolation of the viscosity to lower $T$ as indicated in \bild{LC_speeds}, which preeminently reproduces the measured imbibition speeds.

What are the reasons for this discovery? First of all, the confinement to a cylindrical pore (rather than to a film geometry like in the Couette flow depicted in \bild{LC_visc}) renders the $\eta_3$ viscosity as unfavorable as the $\eta_1$ viscosity. This again clarifies the high stability of the nematic $\eta_2$-configuration in confinement. A cross-over behavior as ascertained for the freely flowing bulk liquid can hence be excluded. Yet, the presmectic divergence of $\eta_2$ is caused by the destruction of precursors of smectic planes for $\vec{n} \,||\,\vec{v}$. From this point of view the suppression of the A phase is not surprising at all but a mere consequence of the overall stabilization of the $\eta_2$-configuration in the nanopore confinement. 

Even the static nanopore-confined liquid crystal shows such modified mesophase behavior \cite{Iannacchione94}. For example the heat capacity anomaly typical of the second-order N-A transition in rod-like liquid crystals immersed in aerogels is absent or greatly broadened \cite{Bellini01, Qian98}. Nuclear magnetic resonance (NMR) measurements revealed the lack of pretransitional smectic layering due to the rough surface of the confining walls \cite{Crawford93}. Furthermore, a systematic study of the influence of the degree of confinement indicates that the N-A transition becomes progressively suppressed with decreasing pore radius whereas the stability range of the nematic phase is increased \cite{Kutnjak03}.

Finally, we will present some results from a more quantitative analysis of the measurements. According to the previously used evaluation method one arrives at slip lengths of $b=(-1.11\pm 0.23)$\,nm for V5 and $b=(-1.54\pm 0.31)$\,nm for V10. Again, a sticking layer boundary condition has to be applied. 

In conclusion the $T$-dependent imbibition measurements of the liquid crystal 8OCB revealed that confinement plays a similar role as an external magnetic field for a spin system: the strong first-order I-N transition is replaced by a weak first-order or continuous paranematic to nematic transition, depending on the strength of the surface orientational field \cite{Stark02}. Based on detailed knowledge of the static (equilibrated) liquid's behavior in the nanopores as deduced from previously conducted birefringence experiments, we were able to procure complementary results with respect to its dynamic (non-equilibrium) behavior. The additional emerging flow velocity and in particular the velocity gradient enhance the paranematic orientational alignment significantly rendering the P-N transition even broader than known from the equilibrium state. Due to the high stabilization of the $\eta_2$-configuration in the N phase the A phase is suppressed and the stability range of the nematic phase is increased. Nevertheless, despite the molecular alignment no indications of velocity slip were found. 

%Apparent velocity slippage at the walls, as was expected to be associated with the molecular alignments in the channels \cite{Heidenreich07}, could not be detected. 

\section{Conclusions}
We have presented a gravimetric study on spontaneous imbibition of a series of molecular liquids as a function of the complexity of their basic building blocks. By thoroughly accounting for the influence of preadsorbed liquid layers this method allowed us to gain insights regarding the flow properties and the phase transition behavior of these systems in extreme spatial confinement. We inferred from our studies an interfacial boundary layer adjacent to the pore walls with a defined thickness whose dynamics are mainly determined by the interaction between liquid and substrate. This manifests itself in terms of a negative velocity slip length for all but a few (more polymer-like) molecular systems investigated here. The flow properties of the pore-condensed molecules in the pore center are, however, remarkable robust, that is bulk-like. This finding of two distinct species (with regard to flow dynamics) is reminiscent of the partitioning of pore condensates found in vapour sorption isotherms (film-condensed versus capillary-condensed state) \cite{Huber99}, of the two species often reported in studies on self-diffusions dynamics (slowed-down or increased self-diffusion versus bulk dynamics) \cite{Baumert2002, Koppensteiner08, Kusmin10a, Kusmin10b}, of the effects of pore condensates on the deformation of the matrix (expansion versus contraction of the pore walls upon change from film-condensed to capillary-condensed state) \cite{Guenther2008}. Last but not least, it reminds of the distinct crystallization behaviour of pore condensates (an amorphous boundary layer versus a crystallized fraction of molecules in the pore center) \cite{Huber99}. In principle, this partitioning, but also the highly stabilized nematic phase of a confined liquid crystal, can be traced to the strong influence of the substrate potential on the liquid or solid layers, respectively, right adjacent to the pore walls \cite{Scheidler00, Scheidler02, Klapp02}. Thus, from a more general perspective, our study highlights that this interaction is not only of importance for an understanding of the equilibrium properties of nanopore-confined condensed matter \cite{Knorr08, Alba06}, but also for the non-equilibrium, flow properties of the molecular assemblies studied here.

For the future the investigation of even more complex molecules and macromolecules such as proteins in mesopores may be envisioned and of interest \cite{Krutyeva2009, Shin07}. Moreover a variation of the substrate species, e.g. a change from porous silica to readily available porous silicon \cite{Gruener08}, porous alumina, porous gold \cite{Liu2008, Kramer04, Biener09} or carbon nanotube \cite{Holt06, Majumder05, Hummer01} membranes, and a modification of the pore wall chemistry, and the resulting change in the liquid/substrate interaction could lead to quite new imbibition phenomenologies. For such studies the concepts and findings presented here may set the stage for a better understanding. 

%Relatively simple gravimetric measurements allow one to explore the rheology of liquids in extreme spatial confinement. The results of the systematic study of the capillary rise of water, hydrocarbons and silicon oils in a network of silica nanopores just presented unambiguously reveal a compartmentation of the pore confined liquid. With high accuracy one perceives an interfacial boundary layer adjacent to the pore walls with a defined thickness whose dynamics are mainly determined by the interaction between {liquid} and substrate. This manifests itself in terms of negative slip lengths. Self-diffusion dynamics (\cite{Baumert2002, Kusmin10a, Kusmin10b}). Dynamical compartimentations another manifestation of the distinction is reminiscent of the findings regarding adsorption (film-condensed and capillary condensed fraction of molecules) \cite{Huber99}, glassy dynamics \cite{Koppensteiner08} and crystallization behavior (amorphous crystalline) even mechanical effects of the matrix (\cite{Guenther2008}).
%
%The physics encountered conventional hydrodynamics, flow physics of porous media with concepts of surface and interfacial physics, in particular also thermodynamical concepts. For the future, we envision experiments on more polymeric systems and ionic liquids, where electrostatic interactions play a dominant role for the future

\section*{Acknowledgments}
It is a great pleasure to thank Andriy Kityk, Klaus Knorr, Dirk Wallacher, and Howard A. Stone for stimulating discussions. Financial support by the DFG under grants Hu850/2 (1-2) within the priority program `Nano- and Microfluidics' is acknowledged.

\section*{References}

\bibliographystyle{iopart-num}
\bibliography{mybib_05042010}

\end{document}